\pdfoutput=1

\documentclass[11pt,twoside,a4paper,cmspaper,final,collab]{cms-tdr}
\def\svnVersion{fcc4125-D}\def\svnDate{2020/06/11}\def\cmsCernNoTag{CERN-EP-2019-266}\def\cmsCernDate{\today}\def\cmsMessage{"Published in the Journal of Instrumentation as \href{http://dx.doi.org/10.1088/1748-0221/15/06/P06009}{\doi{10.1088/1748-0221/15/06/P06009}.}"}

\begin{document}\cmsNoteHeader{PRF-18-003}

\hyphenation{had-ron-i-za-tion}
\hyphenation{cal-or-i-me-ter}
\hyphenation{de-vices}
\RCS$HeadURL$
\RCS$Id$
\newlength\cmsFigWidth
\setlength\cmsFigWidth{0.4\textwidth}

\newcommand{\doserate}{R}
\newcommand{\dosrate}{\ensuremath{R}\xspace}
\newcommand{\ieta}{\ensuremath{i\eta}\xspace}
\newcommand{\iphi}{\ensuremath{i\phi}\xspace}
\newcommand{\doseconst}{\ensuremath{D(R)}\xspace}
\newcommand{\avmu}{\ensuremath{\langle \mu \rangle}\xspace}
\newcommand {\SIPM} {SiPM\xspace}
\newcommand {\insitu} {{in situ}\xspace}
\newcommand {\SCSNeightone} {SCSN$-$81\xspace}
\newcommand {\CoSixty} {$^{60}\mathrm{Co}$\xspace}
\newcommand {\pp}{$\Pp{}\Pp$\xspace}

\cmsNoteHeader{PRF-18-003} 
\title{
Measurements with silicon photomultipliers of dose-rate effects in the radiation damage of 
 plastic scintillator tiles in the CMS hadron endcap calorimeter
}

\date{\today}

\abstract{
Measurements are presented of the reduction of signal output
due to radiation damage for  two types of plastic scintillator tiles
used in the hadron endcap (HE) calorimeter of the CMS detector.
The tiles were exposed to particles produced in proton-proton (\pp) collisions
at the CERN LHC with a center-of-mass energy of  13\TeV,
corresponding to a delivered luminosity of 50\fbinv.
The measurements are based on readout channels of the HE that were instrumented
with silicon photomultipliers, and are derived using data from several sources:
a laser calibration system, a movable radioactive source,
as well as hadrons and muons produced in \pp collisions.
Results from several irradiation campaigns using \CoSixty sources are also discussed.
The damage is presented as a function of dose rate.
Within the range of these measurements, for a fixed dose the damage 
increases with decreasing dose rate. 
}

\hypersetup{
pdfauthor={CMS Collaboration},
pdftitle={Measurements of dose-rate effects in the radiation damage of plastic scintillator tiles using silicon photomultipliers},
pdfsubject={CMS},
pdfkeywords={CMS, Radiation damage to detector materials (solid state); Radiation-hard detectors;
Scintillators and scintillating fibres and light guides; Calorimeters}}

\maketitle

\section{Introduction\label{sec:introduction}}

Because of their versatility and low cost, plastic scintillators are used in the construction of detectors
built for experiments at particle colliders.
They are, however, subject to a reduction in their signal output after irradiation (radiation damage)~\cite{sauli}.
Two of the hadron calorimeters (HCAL) of the CMS detector~\cite{Chatrchyan:2008zzk}
---the hadron barrel (HB)~\cite{Abdullin2008} and the hadron endcap (HE)~\cite{HCALTDR1997}---
at the CERN LHC~\cite{LHC} use tiles constructed from
plastic scintillator  with embedded wavelength shifting (WLS) fibers to produce their signals.
There are also plans to use scintillators in the CMS endcap calorimeters  upgraded for the high-luminosity
LHC runs~\cite{hgcaltdr}.

This paper presents results on the reduction of signal collected from irradiated
scintillator tiles as a function of dose rate \dosrate.
These results provide unique information about radiation damage
at dose rates significantly lower than previously studied.
The HE tiles, described in Sec.~\ref{sec:CMS}, and their associated fibers, were irradiated by
particles produced in \pp collisions at the LHC
during 2017 at a center-of-mass energy of 13\TeV,
corresponding to a delivered luminosity of 50\fbinv.
The \dosrate range is extended by including studies of tiles placed in a moderate-\dosrate
region of the CMS collision hall forward of the HE,
as well as tiles irradiated using external high-dose-rate \CoSixty sources.
The reliability of the measurements is improved by using
tiles that were instrumented before the 2017 data-taking period
with silicon  photomultipliers ({\SIPM}s, also known as Geiger Mode Silicon Avalanche Photodiodes).
The HE tile results are obtained using several complementary methods.
We use a  movable radioactive source that can access all the tiles to compare
their signal output before and after the 2017 data-taking period.
Inclusive energy deposits from \pp collisions
and energy deposits by isolated muons  are also  used to monitor the signal output.
In addition, some of  the HE tiles and the tiles in the moderate-\dosrate region of the collision hall
are studied using a laser calibration system.
The results indicate an \dosrate-dependent effect;
scintillators receiving the same ionizing dose at different dose
rates have different reductions in collected signal.

This study supersedes our previous results~\cite{1748-0221-11-10-T10004}, which were based on
data collected in 2016 using hybrid photodiodes (HPDs)
as the photodetectors. Those photodetectors were subsequently shown to have suffered significant 
response degradation over the course of the running period 
because of damage to the photocathodes by ion feedback~\cite{Dugad}, and not to radiation damage.
In the previous publication~\cite{1748-0221-11-10-T10004}, the reduction of signal output
was attributed solely to radiation damage to the scintillator tiles.

This paper is organized as follows.
In Section~\ref{sec:raddam}, we summarize what is known about radiation damage mechanisms
in plastic scintillators.
In Section~\ref{sec:CMS}, we give a brief description of the CMS
detector, and a more detailed description of the HE calorimeter.
In Section~\ref{sec:results}, we present measurements of radiation damage to the tiles
embedded within the HE.
The calculation of the dose is described, followed by
the results obtained using a laser calibration system to monitor the signal loss,
and using a radioactive source for this purpose.
A parametrization of the {\dosrate} dependence is given.
The signal loss observed in response to hadrons during collisions is studied for consistency
with the laser results, and the signal loss in response to muons is also shown.
In Section~\ref{sec:irradiations}, we present studies of dose-rate effects measured
outside of the CMS detector using irradiation by sources as well as studies
using tiles in the moderate-radiation zone of the CMS collision hall.
In Section~\ref{sec:discussion}, we summarize other relevant information and discuss the dose-rate effects.
Finally, in Section~\ref{sec:summary}, we present a summary and the conclusions of the paper.

\section{Radiation damage mechanisms\label{sec:raddam}}

For the purpose of our studies, we refer to the HCAL tiles as objects consisting of plastic scintillator, a 
WLS fiber, a Tyvek\texttrademark\ wrapping, a clear fiber, and a transducer.
Our estimates, presented below, indicate that the contribution from WLS fibers
 to the overall signal loss is small, and the contributions from clear quartz fibers, Tyvek\texttrademark\ wrappers 
 and the SiPM transducers are negligible.
 Consequently, we believe that our results represent primarily the damage to the scintillator tiles.

Plastic scintillators consist of a plastic substrate, often polystyrene (PS) or polyvinyltoluene (PVT),
into which fluorescent agents (fluors) have been dissolved, usually a primary and a secondary fluor.
When a charged particle traverses the scintillator, the molecules of the substrate are excited.
This excitation can be transferred to the primary fluor via the F{\"o}rster
mechanism~\cite{forster} at primary fluor concentrations above approximately 1\%~\cite{birks}.
The primary fluor transfers the excitation radiatively to the secondary fluor.
For the HCAL tiles made of \SCSNeightone, a  PS-based scintillator from
Kuraray\footnote{Kuraray, Ote Center Building, 1-1-3, Otemachi, Chiyoda-ku, Tokyo 100-8115, Japan},
the absorption maximum of the primary fluor is at
the wavelength of approximately 280\unit{nm}, and the emission is approximately at 320--350\unit{nm}.
The absorption maximum of the secondary fluor corresponds to the emission maximum of the primary fluor, and the
de-excitation of the secondary fluor has a wavelength of maximum emission of approximately 440\unit{nm} (blue light).
This visible light must traverse the scintillator to reach the WLS fiber,
and can be reduced by imperfections in the material (color centers) along its path.

Generally, the scintillator signal output
decreases exponentially with the dose received, as expected for light attenuation due to radiation-induced color centers;
this behavior was also observed in source measurements~\cite{HCALTDR1997}, which were used to design the HCAL optics:
\begin{linenomath}
\begin{equation}
L(d)  = L_0 \exp(-d/D) = L_0 \exp(-d \, \mu),
\label{eqn:exp}
\end{equation}\end{linenomath}
where $L(d)$ is the signal output after receiving a dose $d$, $L_0$ is the signal output before irradiation, $\mu$ is
a function that depends on the dose rate $\doserate$, and $D = 1/\mu$.
When the damage is small compared to measurement uncertainties,
$D$ fluctuates to large positive or negative numbers.
Therefore $\mu$ is used to fit the data and evaluate the uncertainties.
The fitted values of $\mu$ can be averaged over bins of dose rate to improve statistical accuracy.
The \avmu results are used to parametrize the \dosrate dependence
($D$ is shown in some figures of this paper).

The value of {$\mu$}  depends on the materials used in the fabrication of the 
scintillator and on how it is handled
(\eg, if it comes into contact with oils, \etc) prior to and during experimental operations.
Several results have been presented on the dependence on dose 
rate~\cite{Biagtan1996125,34504,Wick1991472,289295,173180,173178,Giokaris1993315,1748-0221-11-10-T10004,gillen}.
In Refs.~\cite{Giokaris1993315}, the authors saw no change in the signal output 
or attenuation length for \SCSNeightone down to dose rates of 
2\unit{Gy/h},  whereas  the authors of Refs.~\cite{Biagtan1996125,34504} saw effects at dose rates between 
10\unit{Gy/h} and 10\unit{kGy/h}. 
A review of the causes of dose-rate effects, and particularly the prominent role played by 
the diffusion of oxygen and polymer oxidation, is given in Section~\ref{sec:discussion}.

Damage to the fluors can occur~\cite{Wick1991472}, but it is generally small~\cite{berlman,173178}.
Damage to the substrate often results in the creation of radicals, conjugated double bonds,
carbonyl species formed by reaction with oxygen, and trapped electrons, and other structures
that can be color centers. Color centers that interfere with the transfer
of light between the primary and secondary fluors reduce the initial light yield.  
Color centers that absorb the light output by the
secondary fluor reduce the absorption length of the light in the scintillator.

\begin{figure}[hbtp]
  \centering
 \includegraphics[width=0.4\textwidth]{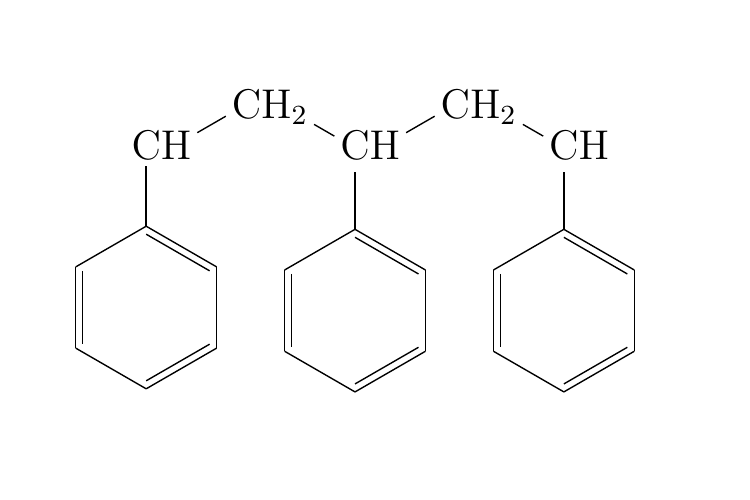}
    \caption{
Polystyrene.
    }
    \label{fig:chemps}
\end{figure}
\begin{figure}[hbtp]
  \centering
 \includegraphics[width=0.8\textwidth]{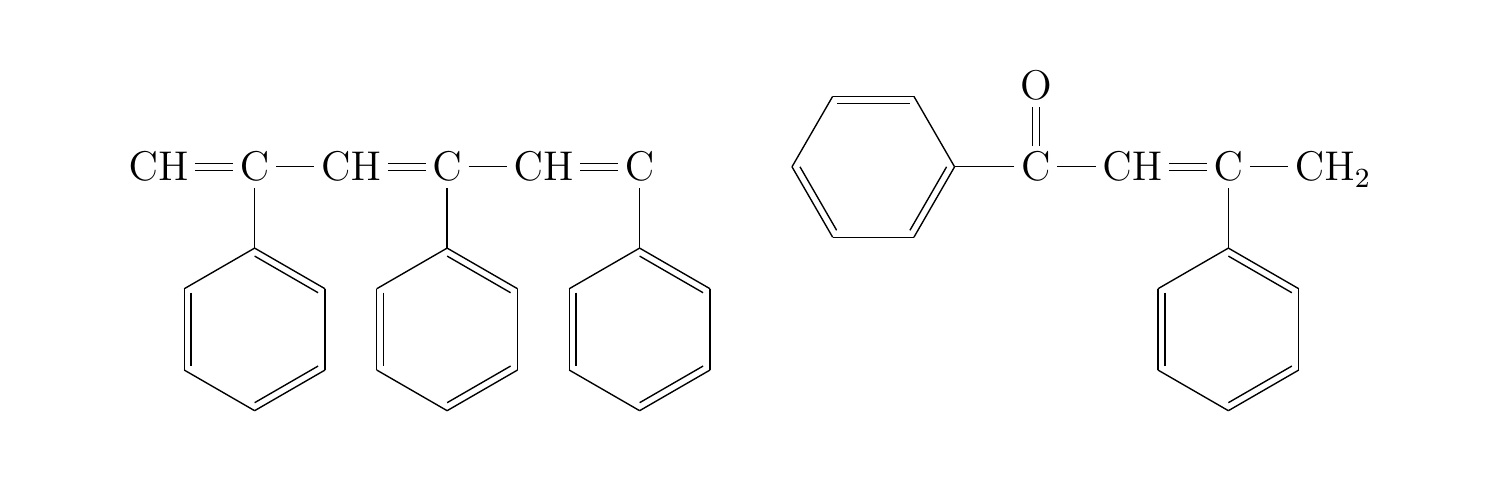}
    \caption{Examples of changes to polystyrene undergoing irradiation.  The change on the right can
only occur in the presence of oxygen.
    }
    \label{fig:color}
\end{figure}

Radicals are produced when chemical bonds in the polymer are broken.
The bonds can re-form on a time scale that depends
on such factors as the density of the radicals and the temperature.
Such damage is called temporary damage, and the re-forming of bonds is known as annealing.
Some products cause permanent changes in the chemical structure.
Figure~\ref{fig:chemps} shows the chemical structure of unirradiated PS.
Figure~\ref{fig:color} shows some of the permanent color centers that can be formed in PS~\cite{cloughrad}.

\section{CMS detector\label{sec:CMS}}

The central feature of the CMS apparatus is a superconducting solenoid of 6\unit{m} internal diameter,
providing a magnetic field of 3.8\unit{T}.
Within the solenoid volume are silicon pixel and strip trackers, a lead tungstate crystal electromagnetic calorimeter (ECAL)
composed of a barrel and two endcap sections, an endcap preshower, and the HB and HE.

The silicon tracker measures charged particles within the pseudorapidity range $\abs{\eta} < 2.5$.
It consists of 1440 silicon pixel and 15\,148 silicon strip detector modules.
Isolated particles of transverse momentum $\pt = 100$\GeV emitted at $\abs{\eta} < 1.4$
have track resolutions of 2.8\% in \pt and 10 (30)\mum in the transverse (longitudinal) impact parameter~\cite{TRK-11-001}.
Muons are measured in the range $\abs{\eta} < 2.4$, with detection planes embedded in the steel flux-return yoke
outside the solenoid that are made using three technologies:
drift tubes, cathode strip chambers, and resistive plate chambers.

A more detailed description of the CMS detector, together with a definition of the coordinate system used
and the relevant kinematic variables, can be found in Ref.~\cite{Chatrchyan:2008zzk}.
A description of the CMS trigger system can be found in Ref.~\cite{Khachatryan:2016bia}.

The scintillator tiles that exhibit damage are located in the HE, which
has 18 layers of active material, denoted layers 0 through 17, over most of its $\eta$ coverage. The zeroth layer of
scintillator uses BC$-$408, a  PVT-based scintillator from the Bicron division of the Saint-Gobain
corporation\footnote{Saint Gobain Corp, Les Miroirs, 18, avenue d'Alsace, 92400 Courbevoie, France},
while the other layers use PS-based \SCSNeightone.
Scintillators based on PVT are brighter than those based on PS.

The scintillator tiles are optically isolated.
They are trapezoidal in shape, and
their faces have a groove shaped like the Greek letter $\sigma$
that holds a 0.94\unit{mm}-diameter Y$-$11 (Kuraray) WLS fiber, mirrored on one end.
The tiles are wrapped in Tyvek\texttrademark.
Clear quartz fibers attached to  the WLS fibers lead to the photodetectors.
Quartz fibers are well known to be radiation hard. In CMS,
we observe small radiation damage to quartz fibers embedded in the Hadron Forward calorimeter,
which is located in a much higher radiation environment than the HE.
  The impact of radiation on the Tyvek\texttrademark\ wrapping is discussed in Sec.~\ref{sec:irradiations}
  and is shown to be negligible.
The tile thickness is 0.9\unit{cm} in layer 0 and 0.37\unit{cm} in the rest of the layers.
When the HE was designed, a thicker and brighter scintillator in layer 0 was
chosen in an attempt to mitigate the noncompensating response of the ECAL to hadrons
and the large amount of dead material installed before the HE for ECAL readout.

\begin{figure}[hbtp]
\centering
\vskip 9mm
 \includegraphics[width=0.9\textwidth]{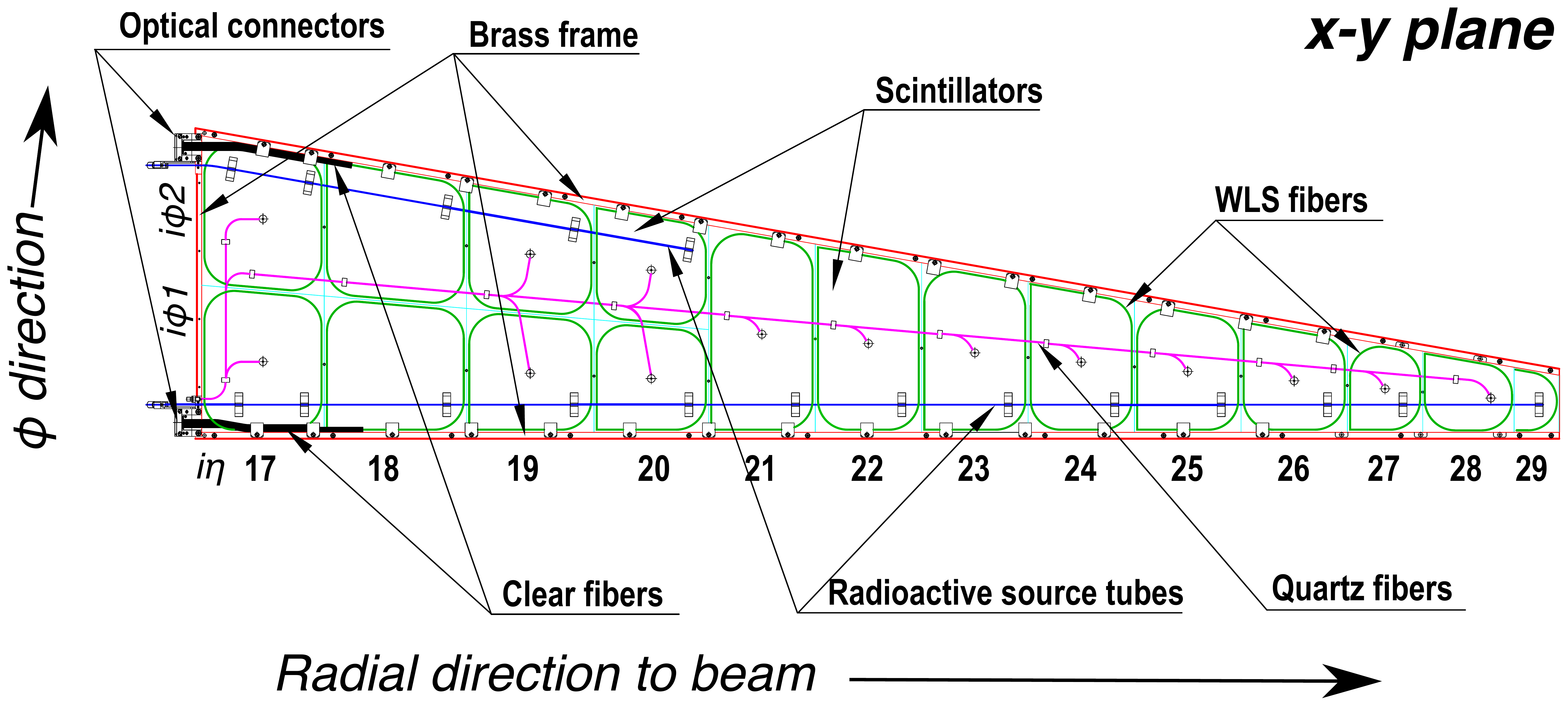}
\vskip 9mm
 \caption{
Details of an HE megatile showing the scintillator tiles, the WLS fibers, and the clear readout fibers.
Also shown are the quartz fibers, which carry the laser light and the tubes through which the radioactive source moves.
In layer 1, the inner size of the megatile is around 7.3\unit{cm}, while the outer size is 38.5\unit{cm}
and the radial extent is 175\unit{cm}.
The sizes (the longer base and the height) of enclosing trapezoids vary between $9.6\unit{cm} \times 12.1\unit{cm}$
for the smallest ($\ieta = 27$), and $13.6\unit{cm} \times 26.5\unit{cm}$ for
the largest ($\ieta = 21$) tile used in this analysis.
}\label{fig:megatile}
\end{figure}

The HE geometry is projective in $\eta$-$\phi$-$z$ space,
where $\phi$ is the azimuth and $z$ is the coordinate along the beam line,
with the origin of the coordinate system positioned at the nominal collision point.
Tiles in successive layers are aligned in a ``tower''.
The towers are labeled using integer indices based on their $\eta$ and $\phi$.
For the HE, the \ieta index ranges from 16 to 29, covering $1.305<\abs{\eta}<3$.
The \iphi index ranges from 1 to 72, with $\iphi = 1$ halfway up the detector and 18 and 19 at its top.
A tower corresponds to the hardware associated with an \ieta-\iphi pair.
The tiles  are mounted as mechanical structures called megatiles, shown in  Fig.~\ref{fig:megatile},
which in the HE are installed in layers perpendicular to the beam direction,
and span the range of 400--550\unit{cm} in $\abs{z}$ and 40--260\unit{cm} in radius, depending on $z$.

\begin{figure}[hbtp]
\centering
\includegraphics[width=0.7\textwidth]{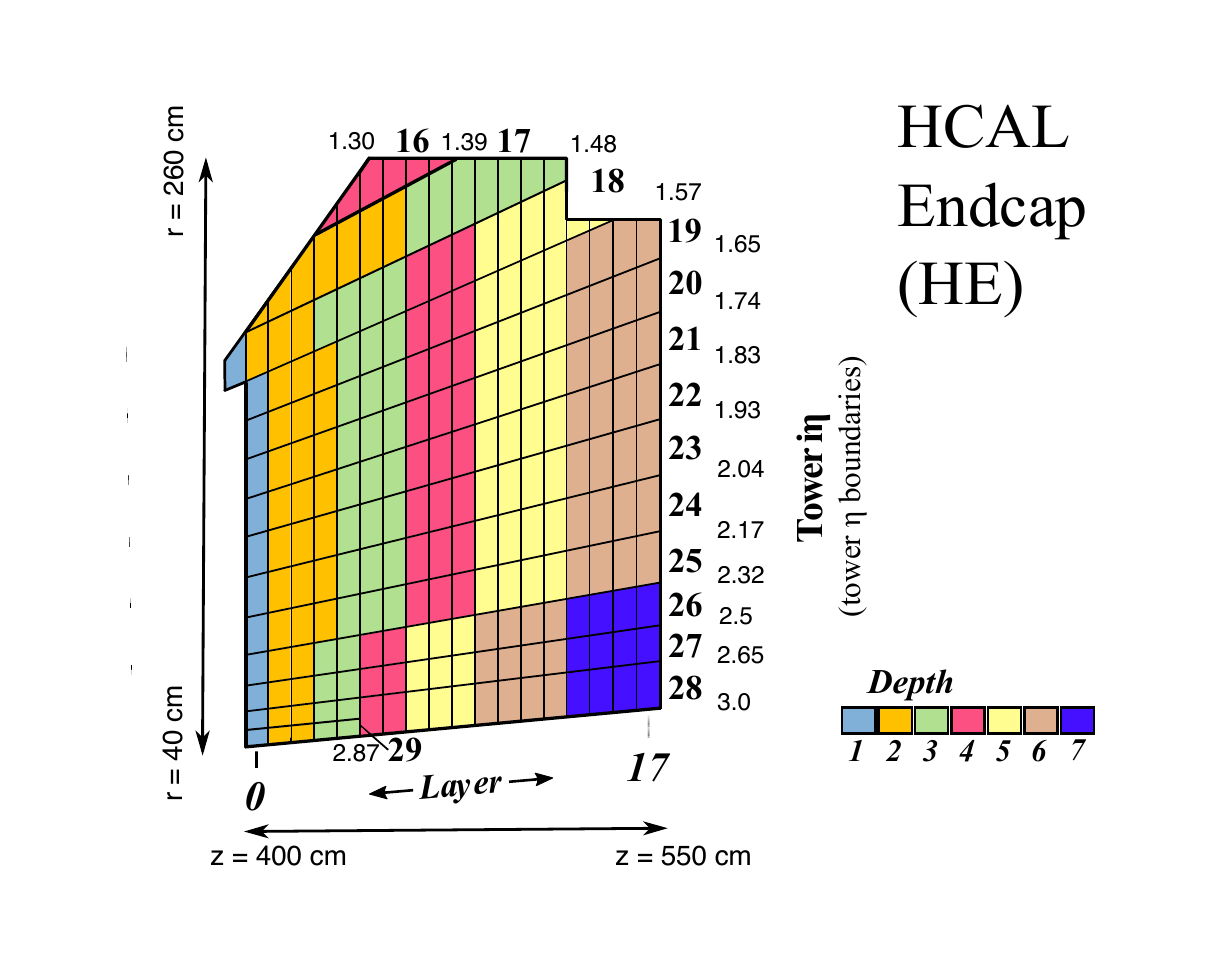}
\caption{
Schematic of the readout segmentation of the HE for channels instrumented with {\SIPM}s.
Scintillator tiles within a tower that have the same color (``depth'') are connected to a single photodetector.
The numbers 0--17 refer to the scintillator layers,
and the numbers 16--29 on the perimeter of the figure denote the \ieta indices of the towers
(the $\eta$ values for the boundaries of the towers are also shown).
}\label{fig:hbseg}
\end{figure}

To limit the number of readout channels, the light from several layers in
a tower is fed to the same photodetector. In the schematic of the HE shown in Fig.~\ref{fig:hbseg},
layers that are fed to a single {\SIPM} have the same color (``depth'').

For data taking prior to 2017, HPDs were used as the HE photodetector~\cite{Cushman}.
For the 2017 data-taking period, tiles in HE towers with
{\iphi} indices of 63--66, corresponding to a $20^\circ$ sector in $\phi$, were read out using  {\SIPM}s.
Our analysis is based on {\iphi}s 63 and 65, because the other {\iphi}s only probed
{\ieta}s below 20 where the radiation damage is too small to be measured reliably.

The HE {\SIPM}s have 2--3 times greater quantum efficiency and
better lifetime response stability than HPDs, no magnetic field sensitivity, require only
medium voltage ($\approx 70\unit{V}$) biasing, have small physical size, and allow the readout of more detector
fibers supporting improved longitudinal segmentation.
 The SiPMs are placed at large radii in the HE, and receive a small radiation dose.
 Test bench measurements of SiPMs irradiated with radioactive sources showed~\cite{Mus2015} that
 the effect of 2017 radiation fluences on the HE SiPM response is negligible.
 Unlike the HPDs~\cite{Dugad}, their gain does not decrease because of light signals received from the tiles.
The primary challenge for \SIPM\ operation is the relatively high dark current resulting from
cumulative radiation damage to the devices \insitu  during future running of
high-luminosity LHC.

The CMS HCAL \SIPM\ devices~\cite{SiPMref} are fabricated by the Hamamatsu
Corporation\footnote{Hamamatsu Corporation, 325-6, Sunayama-cho, Naka-ku, Hamamatsu City, Shizuoka Pref., 430-8587, Japan }.
The approximate device parameters
are 15\unit{$\mu$m} pixel pitch, 4500 pixels per mm$^{2}$, 8\unit{ns} pixel recovery time, and 65\unit{V}
breakdown voltage.
We operate the SiPMs in the Geiger mode at an overvoltage of approximately 3\unit{V},
which corresponds to an operating voltage of about 68\unit{V}.
This value was chosen because it maximizes the signal-to-noise ratio. At this operating voltage, the performance
parameters are approximately 40\unit{fC} per single photoelectron, 12\% pixel crosstalk, and 28\% photon detection efficiency.
Two sizes of circular {\SIPM}s are used: 2.8\unit{mm} diameter devices for depths with
four or fewer scintillator layers and 3.3\unit{mm} devices for the other depths.

A charge-integrating ASIC (QIE)~\cite{qie} is used to read out, digitize,
and encode  the signals from the photodetectors.

Radiation damage to scintillators is sensitive to temperature.
The temperature in the CMS collision hall is about $18^\circ\unit{C}$.

\section{Results from radiation exposure during \texorpdfstring{\pp}{pp} collision data taking\label{sec:results}}

\begin{figure}[hbtp]\centering
 \includegraphics[width=0.7\textwidth]{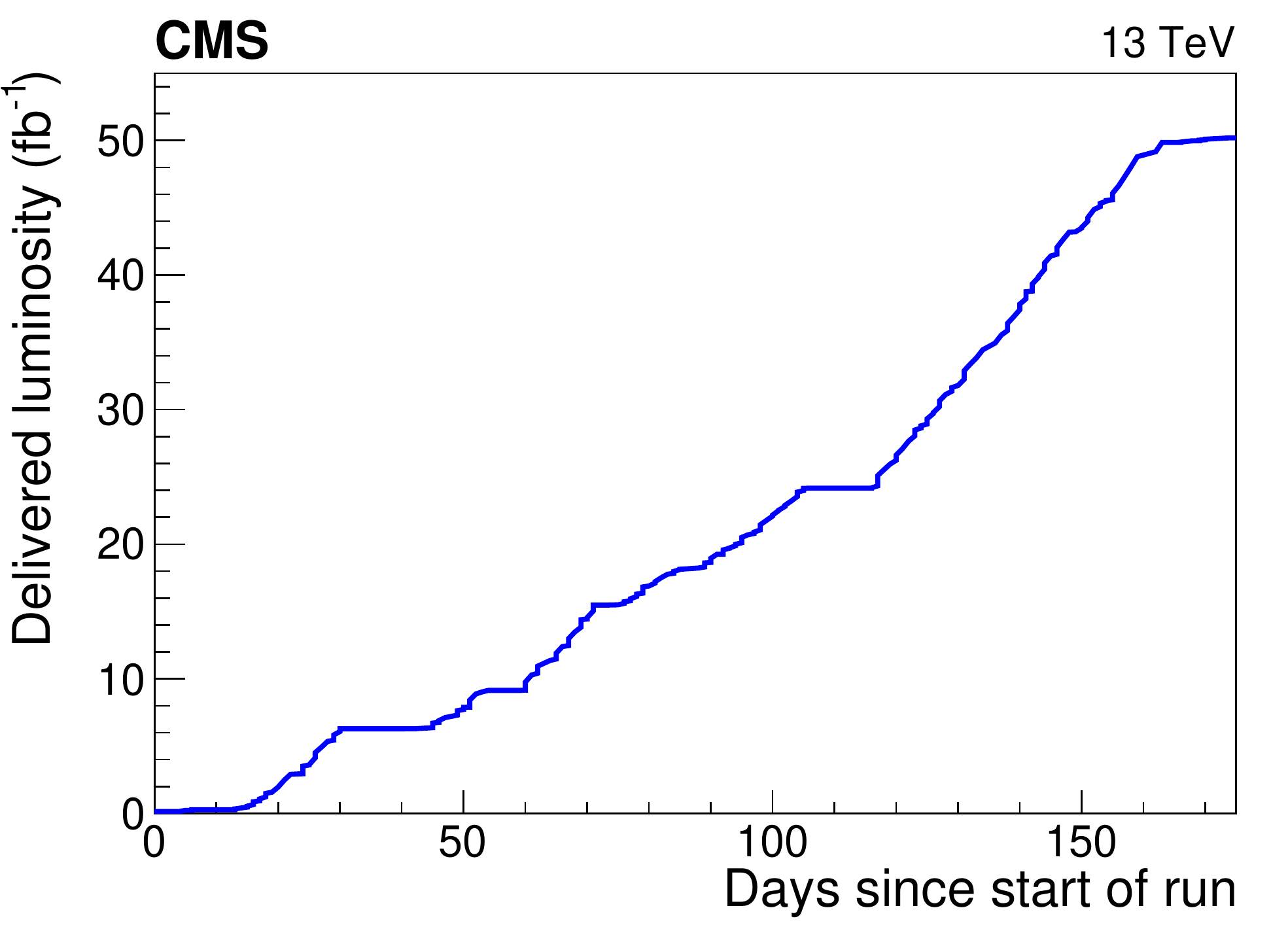}
\vskip 5mm
 \includegraphics[width=0.7\textwidth]{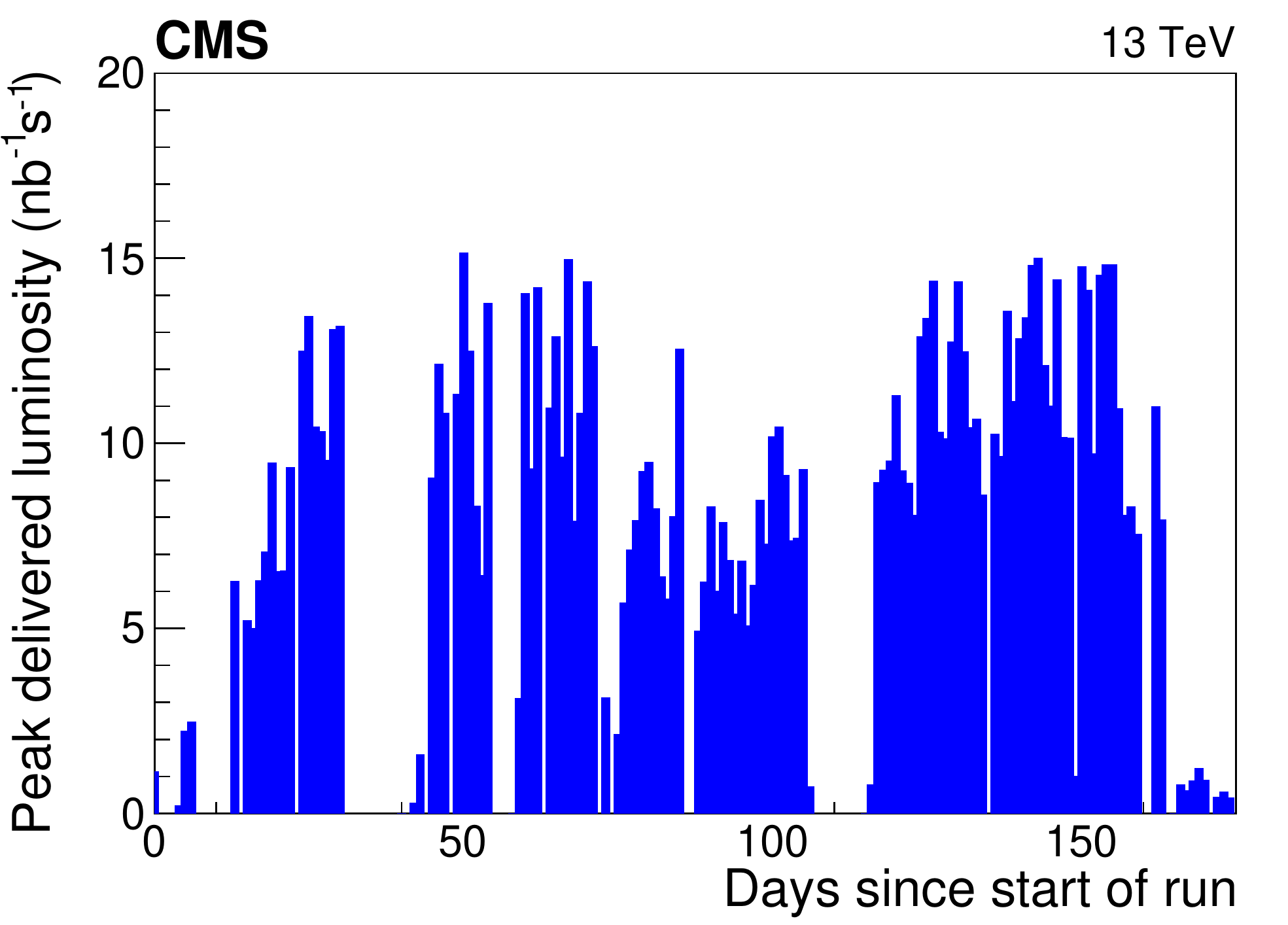}
 \caption{Integrated luminosity delivered to CMS by the LHC in the 2017 \pp data-taking period,
as a function of time (upper) and
maximum  daily (peak) luminosity delivered to CMS in 2017 (lower). Intervals of constant luminosity in the upper plot,
or with no entries in the lower plot, indicate periods with no beam, \eg, technical stops.
}\label{fig:instlum}
\end{figure}

The primary characteristics of the LHC operation relevant
for this analysis are the total delivered luminosity, which determines
the doses received by the tiles, and the average luminosity delivered
per hour, which controls the dose rates.
The  integrated  luminosity delivered as a function of time
as well as the daily maximum instantaneous luminosity
in the CMS interaction region in 2017 are displayed in Fig.~\ref{fig:instlum}.
The daily peak luminosity rose rapidly and then remained
at an approximately constant value  throughout the year.
The mean number of interactions per bunch crossing was about 37.
Multiple interactions present in the recorded beam-beam crossing (event) are
referred to as pileup.

\subsection{Estimation of doses and dose rates in the HE tiles\label{sec:dose}}

For a given luminosity, a tile is subjected
to a dose and dose rate that depend on its location in the detector. The doses
and dose rates vary with pseudorapidity, following
the particle energy density of the \pp collisions, and with depth in
the calorimeter, following the energy deposition profile of
the electromagnetic and hadron showers.

The dose received by each HE scintillator tile per \pp interaction is calculated using simulation
and scaled according to the delivered luminosity.
The calculated doses are verified by \insitu dosimetry.
The peak luminosity versus time was fairly flat during 2017 data taking, indicating
stable running conditions, as shown in Fig.~\ref{fig:instlum} (lower).
We therefore calculate the average integrated luminosity delivered per hour for the whole data-taking period as follows:
for the total of 50\fbinv taken over ${\approx}1670\unit{h}$ of interacting beams we obtain an average integrated luminosity
of 0.03\fbinv/h, with an estimated systematic uncertainty of 5\%.
This value is converted to a dose rate (in Gy/h)
for every HE tile by multiplying the average luminosity per hour by the expected dose per 1\fbinv.

\begin{figure}[hbtp]\centering
 \includegraphics[width=0.7\textwidth]{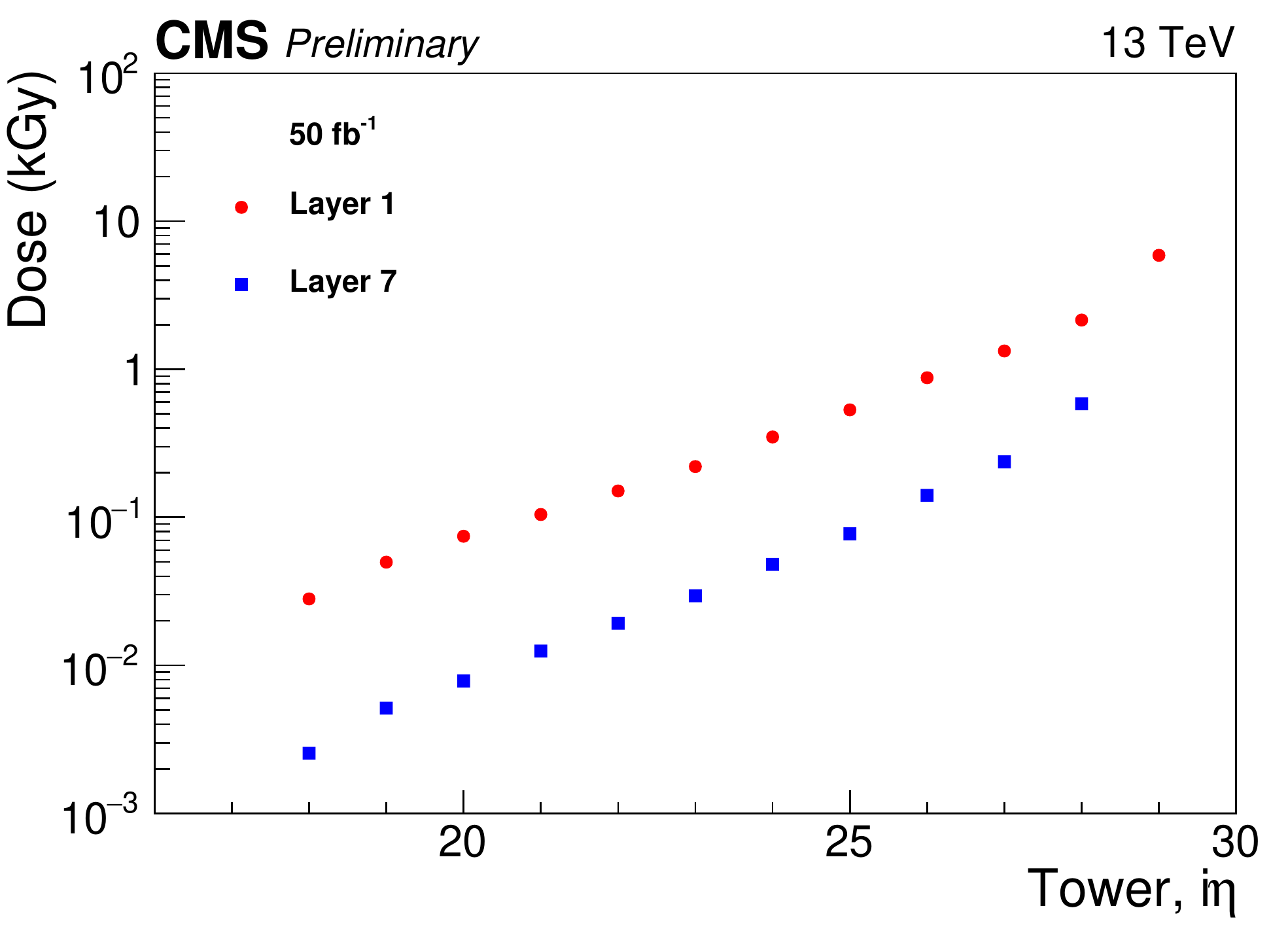}
 \caption{
Doses calculated by FLUKA for the HE tiles in layers 1 and 7
as a function of \ieta for 50\fbinv of LHC running at 13\TeV in 2017.
}\label{fig:dose_vs_ieta}
\end{figure}
\begin{figure}[hbtp]\centering
 \includegraphics[width=0.7\textwidth]{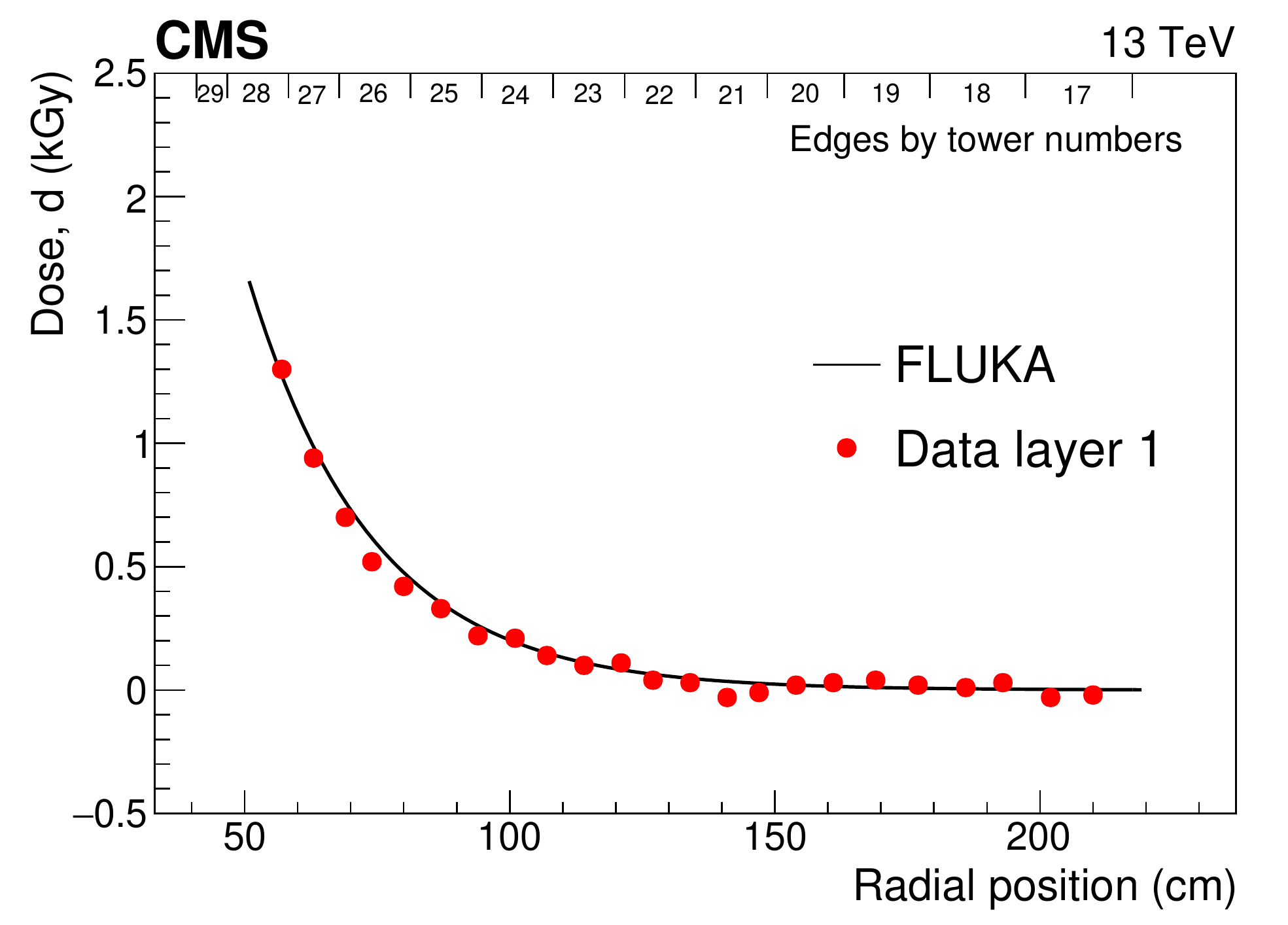}
\vskip 5mm
 \includegraphics[width=0.7\textwidth]{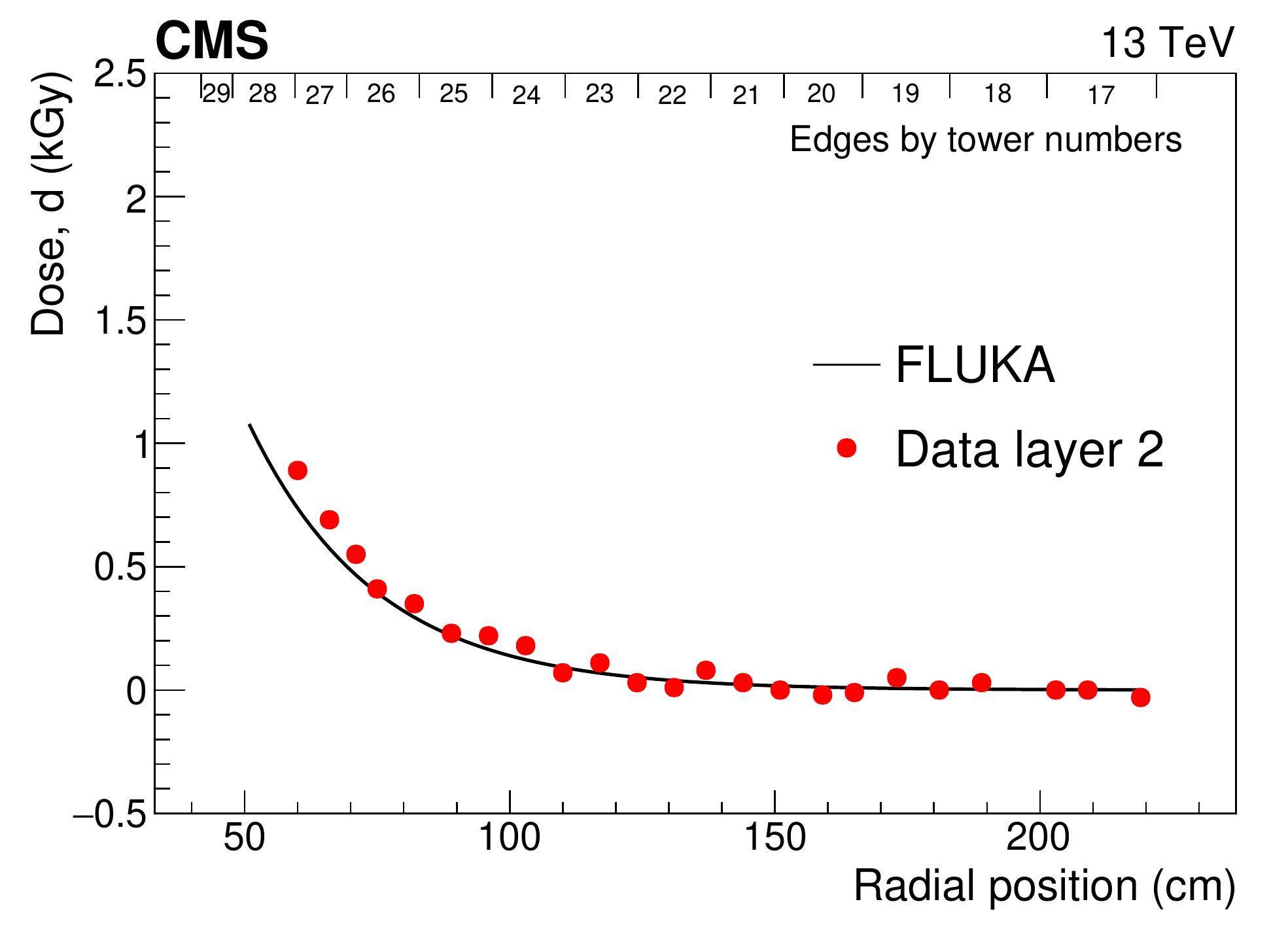}
 \caption{
Comparison of doses for the  2015--2016 data-taking periods calculated using FLUKA and
measured from dosimeter films in layer 1 (upper) and layer 2 (lower), as a function of radial distance from the beam.
Positions of the tile edges in the radial direction are indicated along the tops of the figures.
}\label{fig:smirnov}
\end{figure}

Predictions of the absorbed dose in the HE scintillator layers
are obtained using the Monte Carlo code FLUKA 2011.2c~\cite{fluka1,fluka2}.
The FLUKA predictions for  collisions  use a model that represents the HE in detail,
with brass, Dural\texttrademark ~(Al, Cu, Mg, and Mn), Tyvek\texttrademark,
air, and scintillator layers.
Since the energy loss per unit mass is more than a factor of
two higher for hydrogen than for most other materials, and since  plastic
has a high hydrogen content, the spatial resolution in the simulation is set so that
the dose estimates for tiles does not include regions that are not plastic.
Per 50\fbinv, doses in layer 1  range between  0.03 and 6\unit{kGy}
for {\ieta} of 18 to 29;  for layer 7 they range between 0.003 to 0.7\unit{kGy} for {\ieta} of 18 to 28.
Layers 1 and 7 are located at $z = 410$ and 463\unit{cm}, respectively.
The calculated doses for the 2017 running period for the tiles
in layers 1 and 7 are presented in Fig.~\ref{fig:dose_vs_ieta}.

The calculated doses are verified using measurements with 24 FWT-60 series film dosimeters,
from Far West Technologies\footnote{Far West Technologies, 330 South Kellogg Ave., Suite D, Goleta, CA 93117 USA}
that were installed in the gaps between the absorber and the megatiles in the HE detector
layers 1 and 2 during the 2015 and 2016 data-taking periods, when the detector geometry was essentially the same as in 2017.
The films were measured with a FWT-92D photometer.
The doses were calibrated to water equivalent, which is similar to plastic in terms of density and hydrogen content,
and the uncertainty in the measurements
is estimated to be 3\%. A comparison between the measured and calculated doses
as a function of the distance from the beam line to the film
is given in Fig.~\ref{fig:smirnov}.  Reasonable agreement is seen
for radial distances starting at about 50\unit{cm}, the location
of tower $\ieta = 28$, indicating that FLUKA calculation is accurate
to about 20--30\% for distances 50--120\unit{cm} from the beam, where the largest
radiation damage occurs for the tiles used in this analysis.

The geometry of the detector near towers 28 and 29 is irregular and the dose distribution
difficult to model accurately (due to close proximity to the beam line, beam spray effects, irregular edges of the endcap
preshower and electromagnetic calorimeter, mounting brackets and other construction elements, piping, \etc).
For this reason, data taken for towers 28 and 29 are not included in the fits,
although they are presented in some of the figures below.

\subsection{Results using the laser calibration system\label{sec:laser}}

A laser calibration system is used to monitor the response of the HE
tiles by injecting ultraviolet (UV) light that excites primary fluors
in the scintillator.
It consists of a triggerable excimer laser
and a light distribution system that delivers UV light
(351\unit{nm}) to the scintillator tiles in layers 1 and 7  via quartz fibers.
During the 2017 data-taking period, pulses of laser light were injected
between fills of the accelerator with protons, when there were no collisions.

\begin{figure}[hbtp]\centering
 \includegraphics[width=0.7\textwidth]{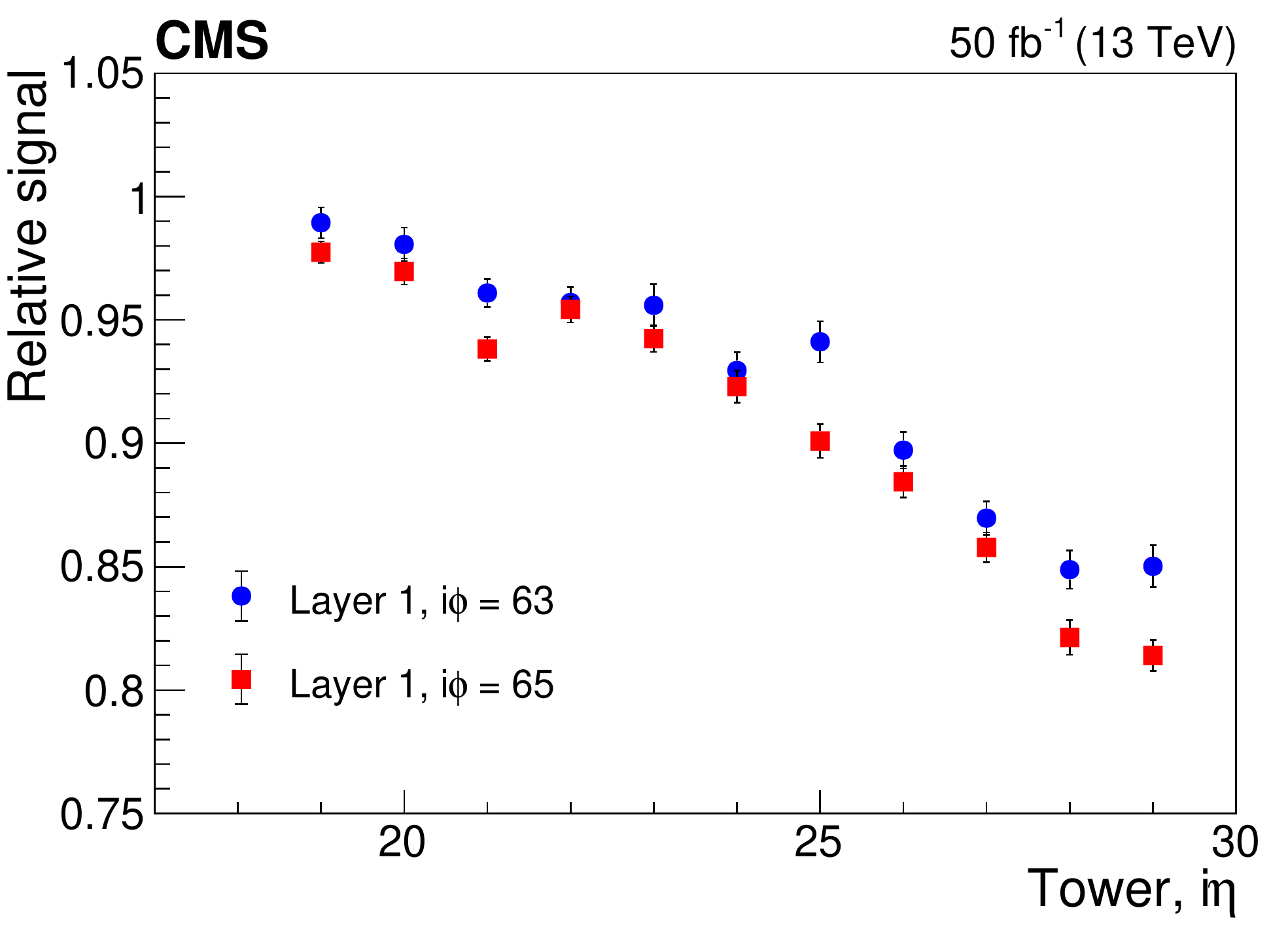}
\vskip 5mm
 \includegraphics[width=0.7\textwidth]{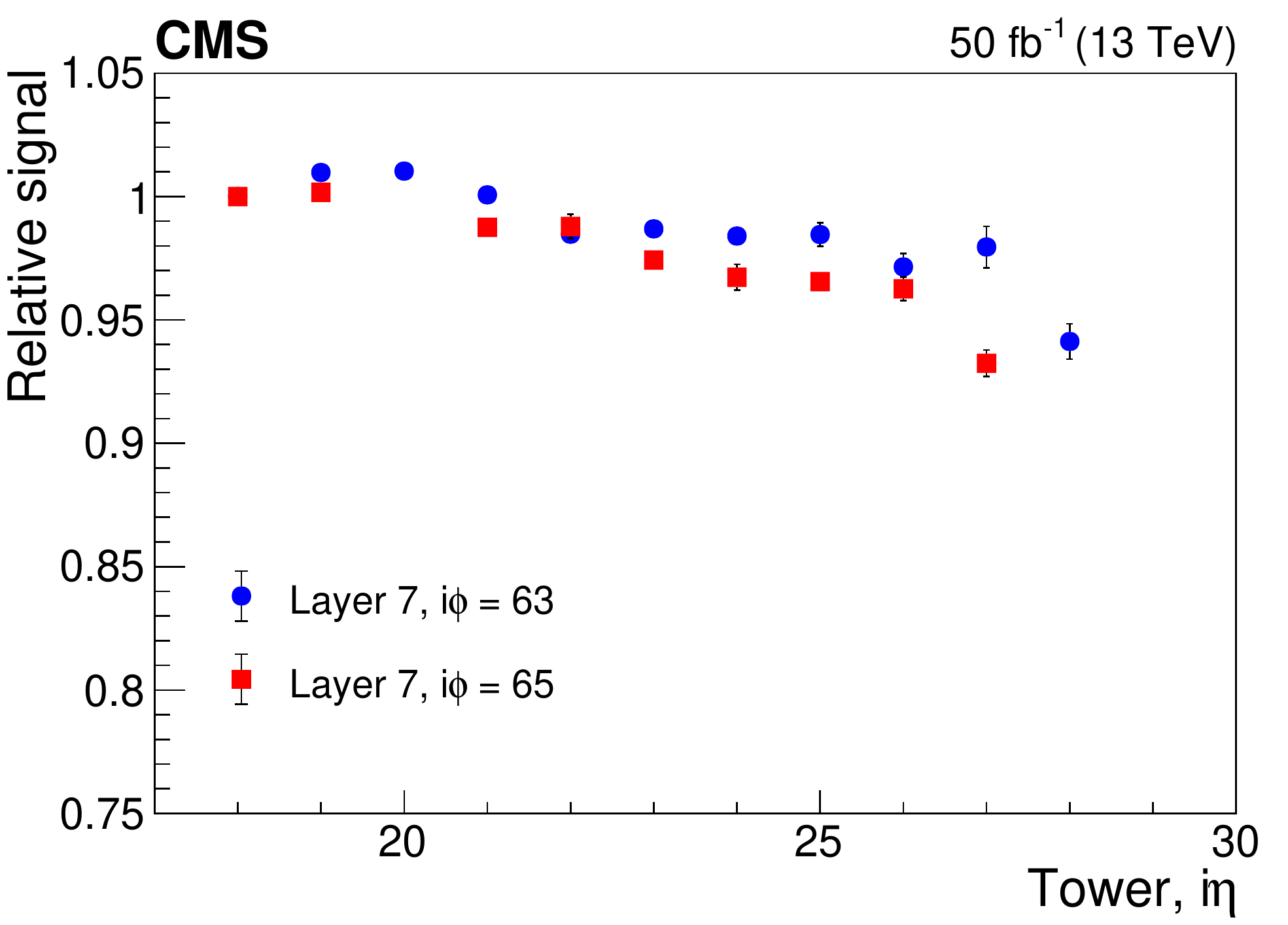}
 \caption{
Signal at the end of the 2017 data-taking period from the HE {\SIPM}s, relative to that at the start,
as measured using the laser calibration system versus {\ieta}
for  \SCSNeightone tiles in layer 1 (upper) and layer 7 (lower).
Only unscaled statistical uncertainties are shown.
The differences between results for the two {\iphi}s indicates unknown systematic uncertainty.
}\label{fig:totlight}
\end{figure}

Laser data were collected throughout the 2017 data-taking period.
Figure~\ref{fig:totlight} shows the signal output for the tiles probed by the laser calibration system
at the end of the 2017 data-taking period relative to that at the start.
Because the intensity of the laser light varied by up to 70\% during 2017,  the signals
are normalized by using signals from tiles at ${\ieta} = 18$ in layer 7, which are expected to have
less than 1\% reduction in signal output. Differences between data for {\iphi}s 63 and 65 are outside
the indicated statistical uncertainties.
These differences contribute to the systematic uncertainties described below.

\begin{figure}[hbtp]\centering
 \includegraphics[width=0.7\textwidth]{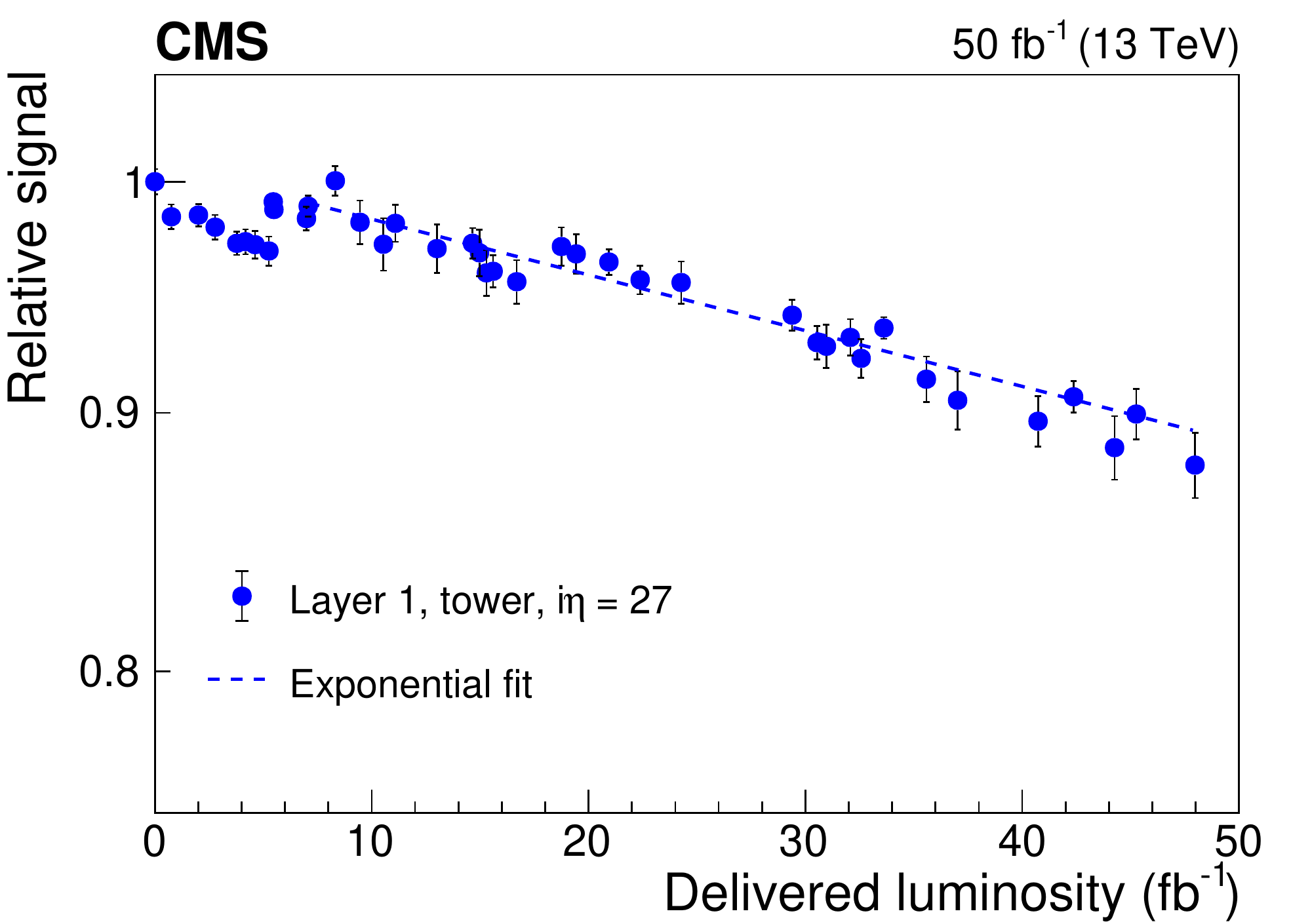}
 \caption{
Relative signal measured using the laser calibration system versus delivered luminosity
for the \SCSNeightone tile in layer 1 with $\ieta = 27$ and $\iphi = 63$.
Scaled statistical uncertainties are shown (see text).
For this tile, the estimated dose at the end of data taking was $d=1.5$\unit{kGy},
and the average dose rate was $\dosrate = 0.89\unit{Gy/h}$.
The dashed line represents a fit to the data to obtain the value of the exponential slope.
Note that the vertical scale is logarithmic  (base 10).
}\label{fig:light}
\end{figure}
\begin{figure}[hbtp]\centering
 \includegraphics[width=0.7\textwidth]{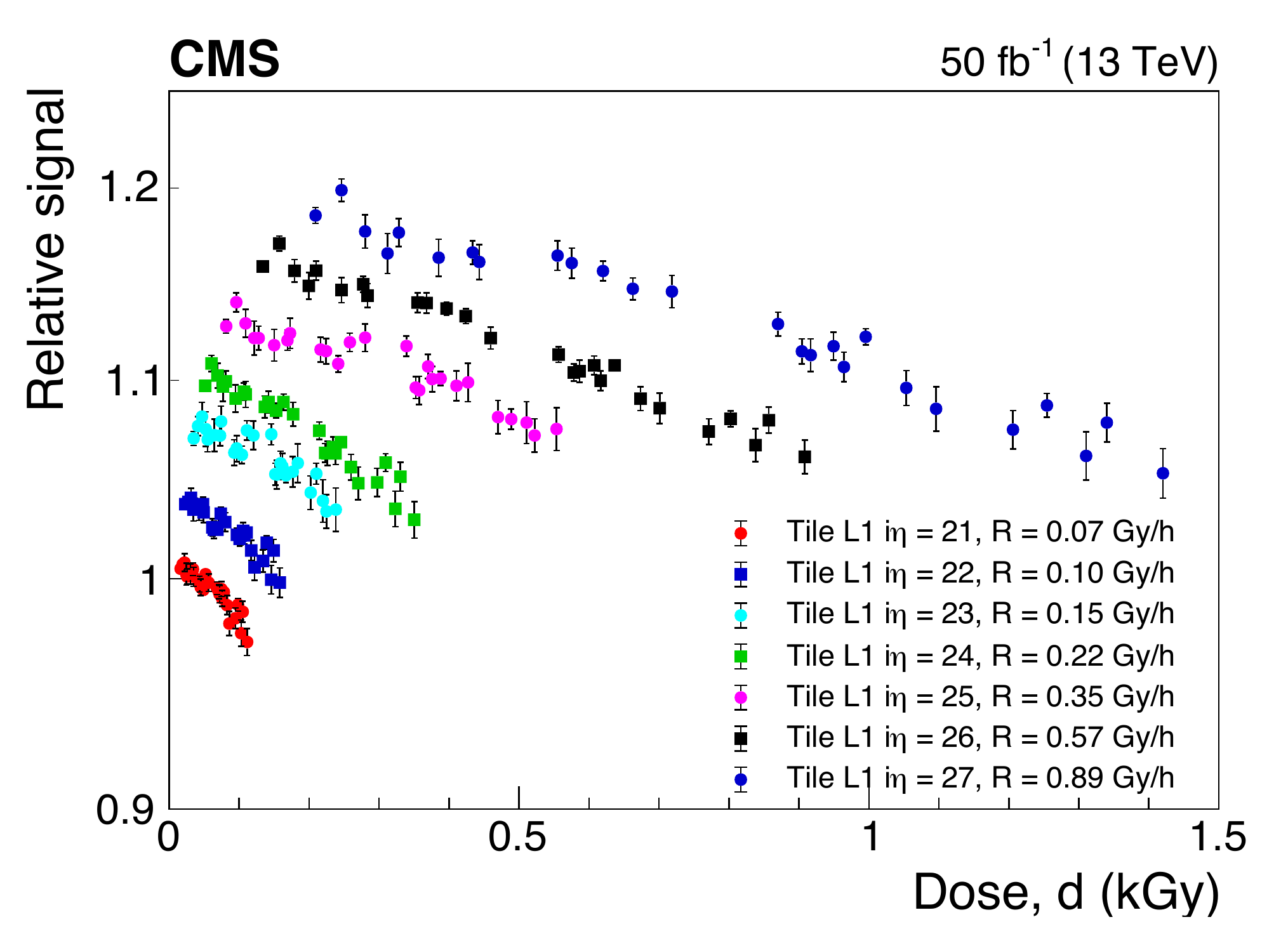}
 \caption{
Relative signal for laser light versus accumulated dose
for the  \SCSNeightone tiles in layer 1 with $\ieta = 21$--27.
The average dose rates are shown for each set of points.
The vertical scale is logarithmic and subsequent sets are shifted  up by a factor of 1.03
relative to the previous set for better visibility.
Each set starts at the dose corresponding to integrated luminosity of 7\fbinv.
Scaled statistical uncertainties are shown (see text).
}\label{fig:lightvsdose}
\end{figure}

The normalized signals from individual channels exhibit an approximately exponential decrease
versus integrated luminosity over most of the data-taking period.
To characterize the behavior of signal loss,
we fit the exponential portions of the normalized signal outputs
with an exponential function of integrated luminosity, as illustrated in Fig.~\ref{fig:light} for one particular tile.
A deviation from the expected exponential behavior is observed during
the first 7\fbinv of data taking. The reason for this effect
is not yet understood and this part of the data is not used in the analysis.
With higher luminosity the effects are clearer so we concentrate on this part of the data.
The statistical uncertainty in the measured mean signal within a single laser run is smaller
than the spread observed when comparing different laser runs taken at similar integrated luminosities.
In consequence, fluctuations  are observed that are larger than expected
based on uncertainties in the mean signal in a single laser
run, indicating the presence of an additional source of scatter. 
We account for these fluctuations by scaling
the uncertainties in individual laser points
to yield a $\chi^2$ per degree of freedom (dof) of one for the exponential fits.
This conservative procedure results in larger estimates of uncertainties in the fit parameters.

Figure~\ref{fig:lightvsdose} presents relative signals versus dose for tiles with $\ieta = 21$--27 in layer 1.
The signals show an exponential decrease (as in Eq.~\ref{eqn:exp}) during periods of stable luminosity,
with slopes that depend on corresponding dose rates.
These results imply that  at a fixed dose the damage to the scintillators increases with decreasing dose rate,
within the range of our measurement.

The values of slopes $\mu$, obtained from the exponential fits,
are averaged in  bins of \dosrate, and converted to
$\doseconst = 1/\langle \mu \rangle$ for comparisons with other measurements of $D$.
Averaging of $\mu$ in bins of dose rate helps to reduce the statistical uncertainties
and extends the range of the measurements to lower values of \dosrate,
especially in the case of source measurements discussed in Section~\ref{sec:source}.
The results for \avmu are discussed in Sec.~\ref{sec:model} and indicate
a dose-rate dependence. A similar dose-rate dependence is also observed without
averaging of $\mu$ in bins of dose rate, but with larger uncertainties in individual points.

We present results for values of {\dosrate} above 0.01\unit{Gy/h}.
The fractional uncertainties in $\mu$ (or $D$) are large for tiles with little damage.
The region $\dosrate > 0.1\unit{Gy/h}$ is well measured with observed signal losses ${>}3$\%.

Various systematic effects have been evaluated. 
In addition to the differences between signals from different {\iphi}s,
we evaluated sensitivities to the variation of the {\ieta} 
choice for normalization, the data range
used for fitting slopes, and the QIE gain setting. Combining these contributions, the overall systematic uncertainty in 
$\mu$ is estimated to be about 25\%.
The measurements are not corrected for the varying sizes of the tiles (see the discussion in
Section~\ref{sec:irradiations}).

\subsection{Results using the radioactive source\label{sec:source}}

Each individual tile in the HCAL is designed to be serviced by a movable \CoSixty
radioactive source using small tubes, which are integrated into the calorimeter.
The \CoSixty source provides photons with energies of 1.17 and 1.33\MeV.
The source is attached to a wire that guides it through the tubes.
All tiles except those in layers 0 and 5, whose tubes have obstructions, can be accessed.
The source moves at approximately 6\unit{cm/s},
and the signal is integrated for 0.1\unit{s} for each measurement.
The resulting signal is used to
monitor the stability of every tile in the HCAL, not just those in layers 1 and 7.
The source data analyzed in this paper were collected during the periods when the LHC did not operate,
both before the 2017 and 2018 data-taking periods.

\begin{figure*}[hbtp]\centering
\includegraphics[width=0.75\textwidth]{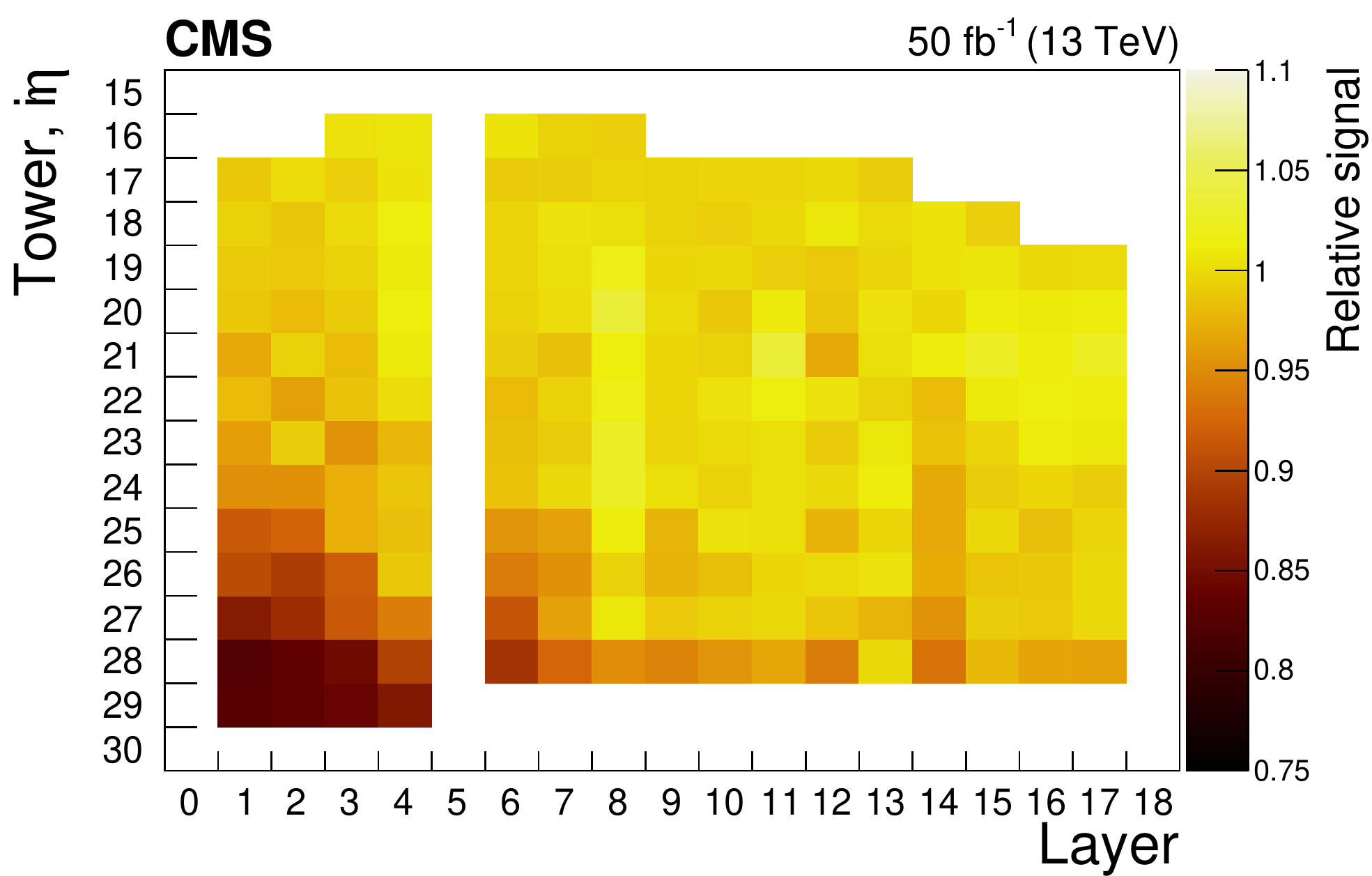}
 \caption{Ratio of \CoSixty source signals observed before and after the 2017 data-taking period,
as a function of \ieta  and layer number  of scintillator  tiles  (\SCSNeightone) in the HE. 
Tubes in layers 0 and 5 have obstructions and cannot be accessed.
}\label{fig:raddammap}
\end{figure*}

The signal strength when the source was far away from a tile is used to estimate
the background.  The measurements of signal output before the 2018 data-taking period are
corrected (divided by  0.886) for the decay of the source since the previous measurements
were made before taking data in 2017. The ratio of the signal
obtained before the 2018 data-taking period to that obtained before the 2017 data-taking period measures the attenuation
of the  signal output due to radiation damage during collisions in 2017, including any post-irradiation
annealing effects.
No additional normalization of signal ratios versus \ieta is required.
Values of the ratio averaged over \iphi
as a function of scintillating tile layer number and tower index {\ieta} are shown in Fig.~\ref{fig:raddammap}.
The signal loss is small for
tiles at large radial distance from the beam and for layers that are deeper in the calorimeter.

At low \dosrate, measurements of signals from individual tiles scatter widely
compared to the expected signal loss, due to the size of the measurement uncertainties.
However, given the large number of tiles measured, a determination of signal loss
can be made even at small  values
of \dosrate assuming that the fluctuations are uncorrelated.
The calculated  $\mu$ values  are averaged in  bins of {\dosrate}
and are displayed in Fig.~\ref{fig:combined_HE_D}.
The uncertainties in \avmu related to the reproducibility of the measurements are included by increasing
the statistical uncertainties by a factor 1.4, which results in the average scatter of points around
the fit being consistent with the scaled uncertainties.
The \avmu values are somewhat lower than, but generally similar to, those from the laser calibration.
The source data represent the damage integrated over
the entire 2017 data-taking period and include an extended annealing
time after the data taking ended.   The analyzed laser data
exclude the first 7\fbinv and any annealing effects
after the end of data taking.

\subsection{Parametrization of laser and source results\label{sec:model}}

\begin{figure*}[hbtp]\centering
 \includegraphics[width=0.7\textwidth]{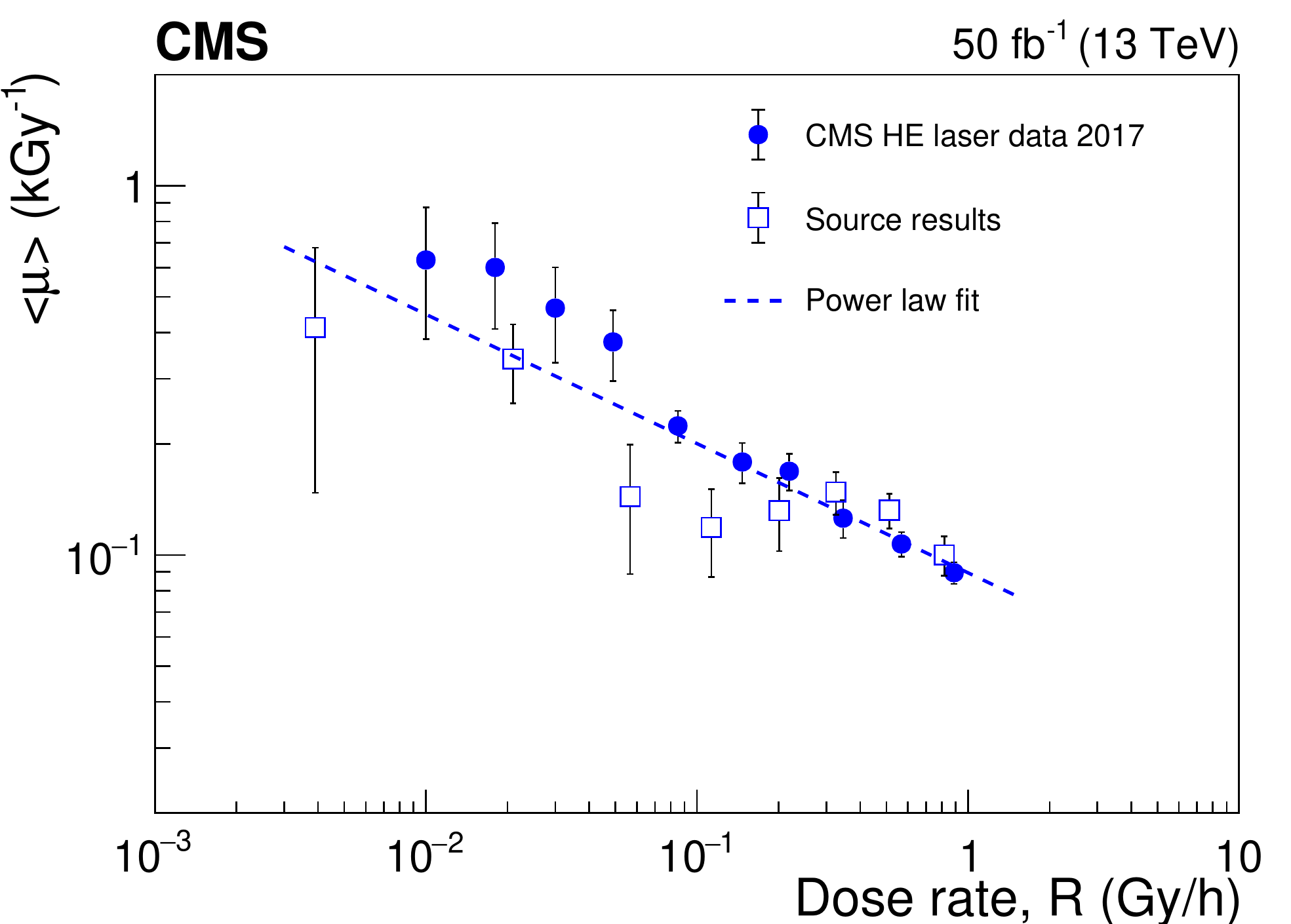}
 \caption{
The value of \avmu  for \SCSNeightone tiles as a function of \dosrate for laser and source
data, parametrized by a power-law behavior, which is shown as a dashed line.
The error bars are dominated by systematic uncertainties.
}\label{fig:combined_HE_D}
\end{figure*}

Figure~\ref{fig:combined_HE_D} summarizes the laser and source \avmu results
for the \SCSNeightone tiles.  
The data are consistent with a power law dependence of \avmu on  {$\doserate$}:
\begin{linenomath}
\begin{equation}
\avmu=1/(\alpha \,  \rho^\beta),
\label{eqn:Dddot3}
\end{equation}
\end{linenomath}
where $\rho = R/R_0$, and the constant $R_0$ can be chosen to minimize the correlation
between parameters $\alpha$ and $\beta$; the fitted value of $\alpha$ depends on the choice of $R_0$.
This form is equivalent to $D=\alpha \,  \rho^\beta$.
The value of $R_0 = 0.32$\unit{Gy/h} is chosen for the fits below
so that the correlation between parameters $\alpha$ and $\beta$ becomes negligible.
The dashed line shown in Fig.~\ref{fig:combined_HE_D}
is the result of a power-law fit to both sets of data assuming all uncertainties are uncorrelated.
The corresponding model parameters are
$\alpha=7.5\pm 0.3$\unit{kGy}  and $\beta=0.35\pm 0.03$ when
\avmu is in \unit{kGy$^{-1}$} and {\dosrate} is in units of \unit{Gy/h}.
The fit $\chi^2$/dof is 1.2.
A fit to the laser data alone yields $\alpha=7.3\pm 0.3$\unit{kGy}
and $\beta=0.43\pm 0.04$, with a $\chi^2$/dof of 0.4.
A fit to source data alone gives $\alpha=7.6\pm 0.5$\unit{kGy}
and $\beta=0.21\pm 0.06$, with a $\chi^2$/dof of 1.1.
The fit to the laser data is inconsistent with no dose-rate effect.
The fit to the source data by itself shows a smaller dose-rate effect, and
is inconsistent with no dose-rate effect at the 3.5 standard deviation level.
For the parameter $\beta$, which measures the dose-rate dependence,
the difference between the results from the laser and source fits is $0.22\pm 0.08$ (2.7 standard deviation).
The tension between laser and source results may be a fluctuation.
Since the \avmu values from the source data tend to be lower than
those from the laser data, additional annealing between the end of
\pp collisions and the source scan is a possibility.
Annealing reduces damage and therefore decreases $\mu$.
A future source measurement of the HE and a measurement of annealing effects
using post data-taking laser runs
would help to reduce this uncertainty.

The systematic uncertainty in parameter $\alpha$ is assumed to be the same as the 25\% systematic uncertainty in $\mu$,
discussed in Sec.~\ref{sec:laser}, assuming a 100\% correlation between the measurements.
For the parameter $\beta$, the spread of fit results between the laser and source data
indicates systematic effects of the order of 0.1, when varying the range in {\dosrate} used in the fit.

The parametrization of our results should be used with care. It is valid for the
decrease in signal output for a system consisting of scintillators,
wavelength shifting fibers, and clear fibers made from the same materials we used, and constructed
in the CMS tile geometry, when irradiated in the environment of the CMS collision hall.
Kuraray has indicated that the current Y$-$11 fiber is not the same as past versions.
The parameter values are not generally applicable for other scintillator systems.
Extrapolation of the power law above a dose rate of ${\approx}10$\unit{Gy/h}  is not expected to be valid.
As discussed in Sec.~\ref{sec:discussion}, at {\dosrate} of approximately 10\unit{Gy/h},
oxygen will no longer permeate the entire  tile~\cite{cloughPS,Wick1991472}.
Radical creation and termination is different in regions with and without oxygen.

\subsection{Cross-checks with inclusive hadrons\label{sec:hadrons}}

An additional method of measuring the effects of irradiation on the
tiles is based on the 2017 collision data.
Radiation damage is studied using observed energy depositions from hadrons produced in \pp collisions.
The energy distribution is measured for 25 subsamples
distributed uniformly in delivered luminosity over the entire 2017 data-taking period.
For each data-taking period $n$, the ratio of average energy relative to that of period 1,
\begin{linenomath}
\begin{equation}
F_{\text{meas}}(n) = \frac{E_{\text{ave}}(n)}{E_{\text{ave}}(1)},
\label{eq_raddam}
\end{equation}\end{linenomath}
serves as a measure of the radiation damage, where
$E_{\text{ave}}$ is the average signal measured in all readout channels
with the same values of {\ieta} and depth; the average is
calculated from the sum of signals above the  threshold of $E_{\mathrm{min}} = 0.5\gev$.

\begin{figure}[hbtp]\centering
\includegraphics[width=0.7\textwidth]{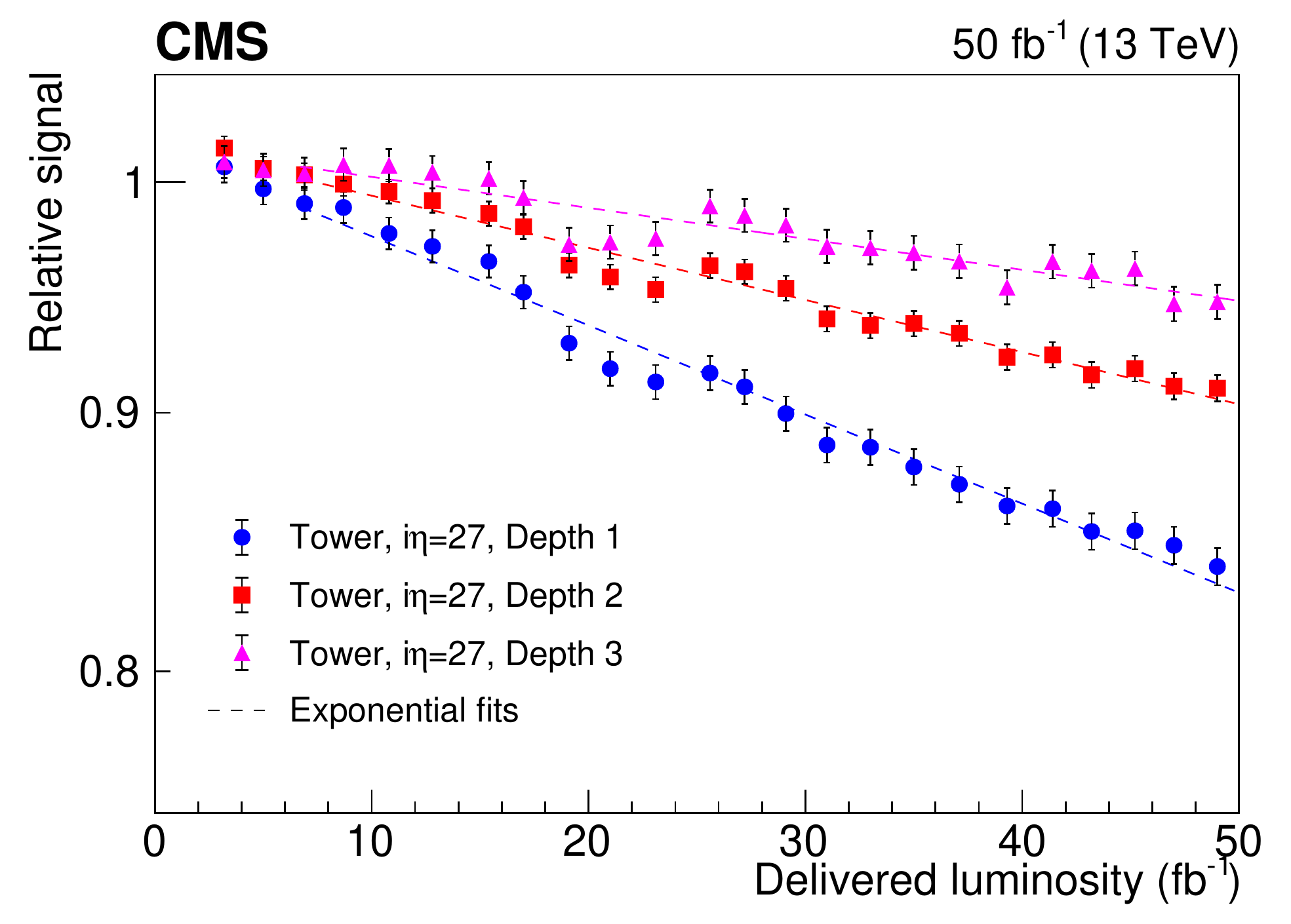}
\vskip 9mm
\includegraphics[width=0.75\textwidth]{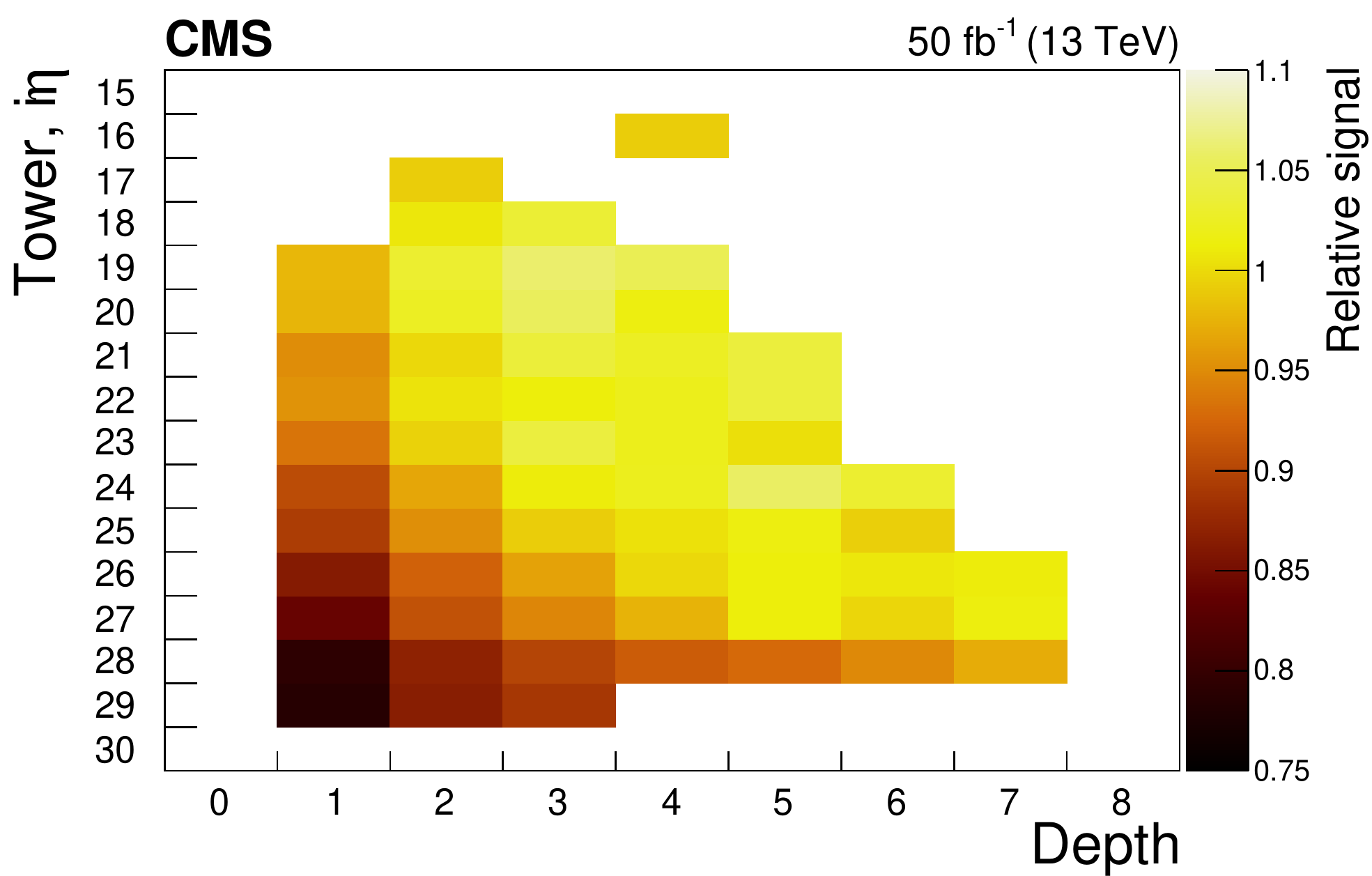}
\caption{
Upper: Relative signal $F$ for $\ieta = 27$ in depths 1, 2, and 3 versus delivered
luminosity using the in situ ``inclusive'' method; the dashed lines show the results of fits with an exponential
function, after excluding the first 7\fbinv of data,
as was done in the laser data analysis (Sec.~\ref{sec:laser}).
For the tile in depth 1 (\ie, layer 0), the estimated dose at the end of data taking was  $d=1.5$\unit{kGy}
and the average dose rate was $\dosrate = 0.89\unit{Gy/h}$.
Lower: Relative signal $F$ for towers with $\ieta = 16$--29 at different depths measured after 50\fbinv of
delivered luminosity; only results with a relative uncertainty of 3\% or lower on measured values of $F$ are shown.
Tiles in depth 1 are made of BC$-$408 and tiles in other depths are \SCSNeightone.
 }
\label{fig:haddepth}
\end{figure}

The energy comparison requires a selection of events that is both independent of
the HCAL and selects a well-defined set of hard interactions that is stable throughout the period under study.
This is fulfilled by utilizing events satisfying a dimuon trigger. The energy ratio is studied as a
function of the average number of interactions per bunch crossing, $n_{\mathrm{PU}}$,
to take into account the difference in the pileup structure between the periods.
The number $n_{\mathrm{PU}}$ is estimated from the instantaneous luminosity.

For each value of {\ieta} and depth, the pileup dependence of $F_{\text{meas}}$
is eliminated by fitting it versus $n_{\mathrm{PU}}$ with a linear function.
The fits are performed in the range $20 < n_{\mathrm{PU}} < 50$
and the values of $F_{\text{meas}}$ are extracted at $n_{\mathrm{PU}}=35$.

The ratio $F_{\text{meas}}(n)$ at  $n_{\mathrm{PU}}=35$ is observed to depend
on the energy threshold  $E_{\mathrm{min}}$.
Both the numerator and denominator of $F_{\text{meas}}(n)$ are sums
of energies of those individual channels that are above the threshold $E_{\mathrm{min}}$.
In the presence of radiation damage the ratio $F_{\text{meas}}(n)$
will typically be smaller than the ratio $F(n)$ that would be
obtained were the threshold not present.
The higher the $E_{\mathrm{min}}$ threshold, the larger the discrepancy.
To correct for this, a calibration is performed as follows.
Using data from the first subsample, we multiply the energies
contributing to the numerator by scale factors that represent hypothetical
signal losses due to radiation damage, but we leave the denominator unchanged.

The values of the scale factors are varied in the range observed in the data,
and for each scale factor $F'$ a value $F'_{\text{meas}}$ is extracted using the method described above.
A linear relationship between $F'$ and $F'_{\text{meas}}$ is found, which is used to correct
the measured values of $F_{\text{meas}}(n)$ to obtain the corresponding $F(n)$.
The magnitude of this correction depends on {\ieta} and depth,
and typically amounts to no more than 20\% of the measured signal loss fraction ($1-F_{\text{meas}}(n)$).

\begin{figure}[hbtp]\centering
\includegraphics[width=0.7\textwidth]{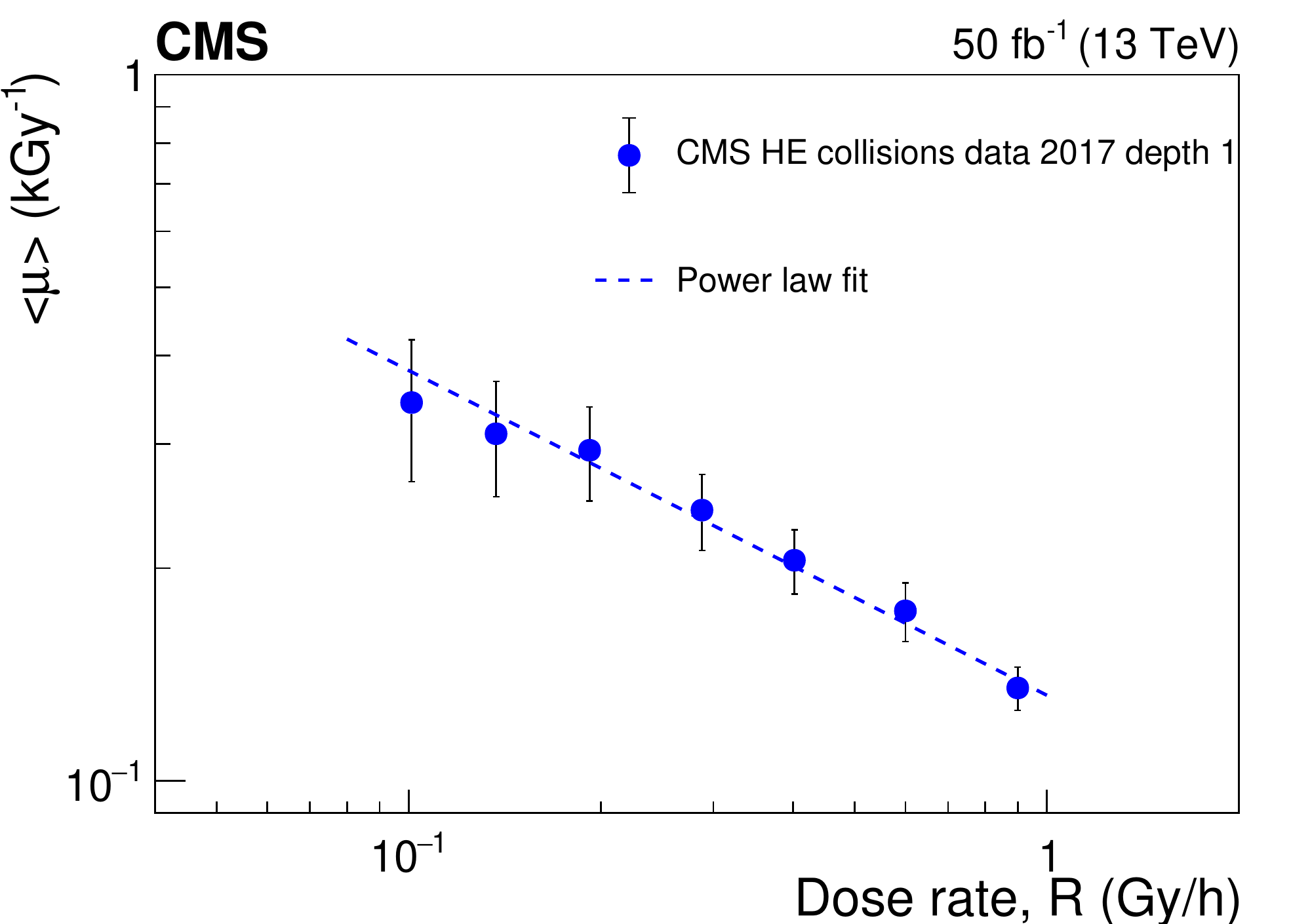}
\caption{
The value of \avmu as a function of \dosrate for in situ collision
data in depth 1 (BC$-$408), parametrized by a power law behavior, which is shown as a dashed line.
}
\label{fig:insitu_lzero}
\end{figure}

The corrected signal fractions $F$ measured for the channels in the first three depths of
${\ieta} = 27$  are shown in Fig.~\ref{fig:haddepth} (upper), as a function of delivered luminosity.
The error bars include a systematic uncertainty of ${<}1$\%, which
results in fit $\chi^2$/dof of around one.
A decrease of $F$ with delivered luminosity is clearly seen.
A small shift of points near 20\fbinv is believed to be due to
residual luminosity calibration uncertainty during this period.
Figure~\ref{fig:haddepth} (lower) presents the values of $F$
averaged over {\iphi} as a function of {\ieta} and depth after 50\fbinv,
showing a decrease of $F$ with increasing {\ieta} and decreasing depth.
The behavior is consistent with that shown for individual tiles  observed by the moving source
for all the tiles of the HE, albeit with an increased granularity due to a readout in depths and not layers.

Depth 1 consists of a single layer (layer 0) and thus its tiles have well-defined doses
and dose rates. Using the same procedure as for the laser data,
these data can therefore be converted to \avmu versus \dosrate.  The results are
shown in Fig.~\ref{fig:insitu_lzero}. The parameters of the power-law fit are
$\alpha = 5.4\pm 0.1$\unit{kGy} and $\beta = 0.46\pm 0.04$,
with a $\chi^2$/dof of 0.5, for $R_0 = 0.48$\unit{Gy/h}.
The fit to the layer~0 in situ data is inconsistent with no dose-rate effect.
The layer 0 tiles are constructed from PVT instead of PS, and hence their behavior
can differ from that of PS-based tiles previously discussed.
Nonetheless, the value of $\beta$, which parametrizes the dose-rate dependence, is similar to that from the laser measurements.
At a given dose rate, the values of \avmu are larger (and the value of the $\alpha$ parameter
is smaller) for this PVT-based material, indicating more damage than for the PS-based tiles.

\subsection{Cross-checks using isolated muons\label{sec:muon}}

The most probable energy deposition by a muon can also
be used to estimate the amount of radiation damage.  The acceptance of
the tracker and of the muon system limits this measurement to portions of the HE where
the damage is measured to be small.

\begin{figure*}[hbtp]\centering
 \includegraphics[width=0.7\textwidth]{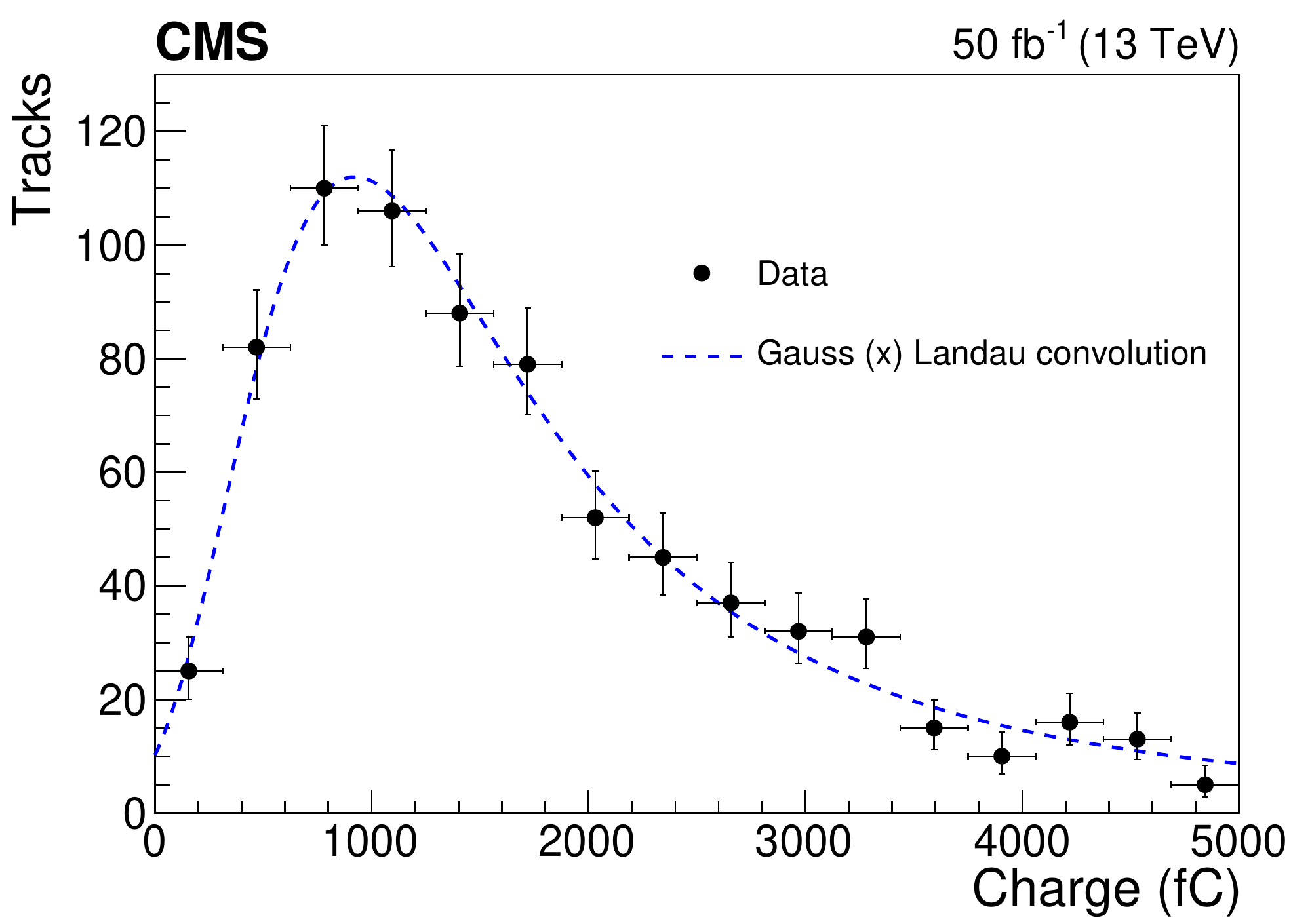}
 \caption{Fit to the charge distribution in an HE tower $\ieta = 26$ depth 1 (BC$-$408) due to an 
isolated muon from one of the event samples of 2017 data.
}\label{fig:muonlandau}
\end{figure*}

The trajectories of forward isolated muon candidates with $\pt>20$\GeV are propagated
to the calorimeter surface to determine which tower they will traverse.
The data-taking period is divided into subsamples.
For each,  a Landau distribution convolved with a Gaussian resolution function is fitted
to the charge distribution from the tower to obtain the most probable value (MPV) of deposited charge.
A typical spectrum, including the fit, is shown in Fig.~\ref{fig:muonlandau}.

\begin{figure}[hbtp]\centering
 \includegraphics[width=0.7\textwidth]{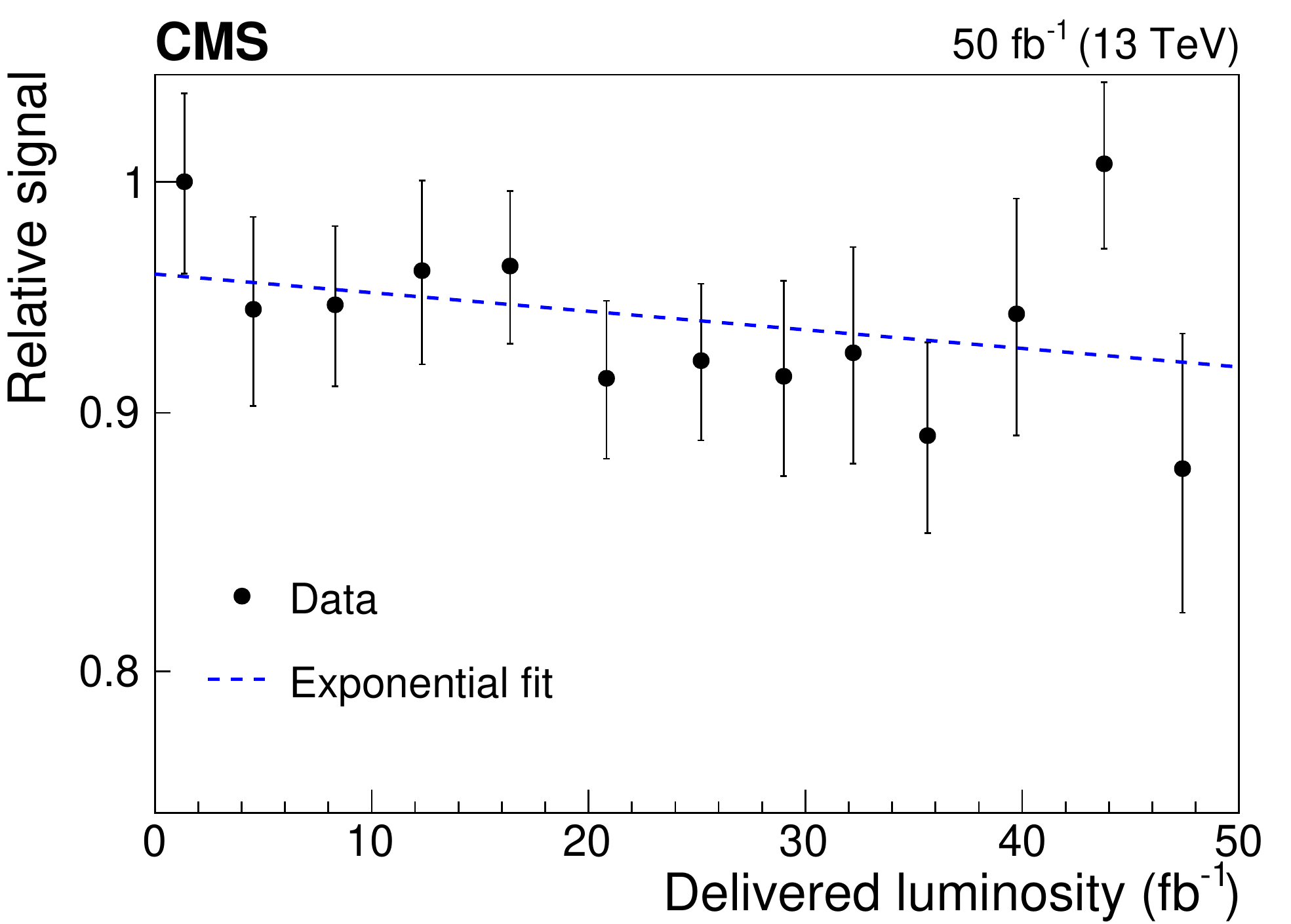}
 \caption{
Relative muon signal in an HE tower with $\ieta = 26$ and depth 1   (BC$-$408)
versus  delivered  luminosity. 
The dashed line shown on the figure is the result of a fit to an exponential distribution.
}\label{fig:muonenergy}
\end{figure}

Because of pileup contributions to the measured signal,
the isolated muon analysis uses events with a similar number of reconstructed vertices
(the range 20--25 was used).
The ratio of the MPV plotted as a function of delivered luminosity to that of the first
subsample for ${\ieta} = 26$ depth 1 is shown in Fig.~\ref{fig:muonenergy}.

Only the towers at shallow depths and large \ieta values are damaged sufficiently
to detect the losses due to radiation damage in 2017 using this technique.
Currently, this measurement is not competitive with other results for these towers.
Upgrades for the CMS detector planned for future operations will have a tracking system
with a larger $\eta$ acceptance, extending the usefulness of this technique.
Monitoring of calorimeter signals with muons has been tried for the first time using the 2017 data.
It is important to develop this technique further for use in future operation.

\section{High-dose-rate results using sources\label{sec:irradiations}}

The CMS laser data monitor the HE tile performance for {\dosrate} only up to about
2\unit{Gy/h} (see Fig.~\ref{fig:combined_HE_D}).
Intense radioactive sources are used to irradiate plastic scintillator tiles and obtain data
at higher {\dosrate},  up to 1\unit{kGy/h}.
To look at {\dosrate}-dependent effects and to avoid bias from other factors, such as tile geometry or chemical composition,
only results from $10\unit{cm} \times 10\unit{cm} \times 0.37\unit{cm}$ \SCSNeightone scintillator tiles read out 
with WLS fibers are reported here, unless noted otherwise.
Although temporary damage is small for tiles irradiated in the HE, it is larger at the {\dosrate} values above
100\unit{Gy/h}. The values reported in this section
reflect the permanent damage to the scintillator tiles remaining after annealing.
This was ensured through observation of the signal output versus time.

Some of the data were taken at facilities with \CoSixty gamma sources,
located at the Kharkov Institute of Physics and Technology (KIPT),
National Research Nuclear University MEPhI,
Goddard Space Flight Center, Argonne National Laboratory (ANL), the Michigan Memorial Phoenix Project,
the National Institute of Standards and Technology in Gaithersburg MD, and at the University of Maryland (UMD).
We also include a measurement from irradiation using an electron beam
at Florida State University (FSU), described in Ref.~\cite{HAGOPIAN}.
For these measurements, some tiles had a fiber with
a slightly smaller diameter, and a more recent formulation of Y$-$11 fiber
from Kuraray than that used for the HE construction.  The machining
of the grooves in the tiles was also performed by different machinists using
different toolings, and different machining rates.
The temperatures of the tiles during the various irradiations are not known precisely,
hence the processes affecting the annealing of radicals may differ somewhat.

\begin{figure}[hbtp]\centering
 \includegraphics[width=0.7\textwidth]{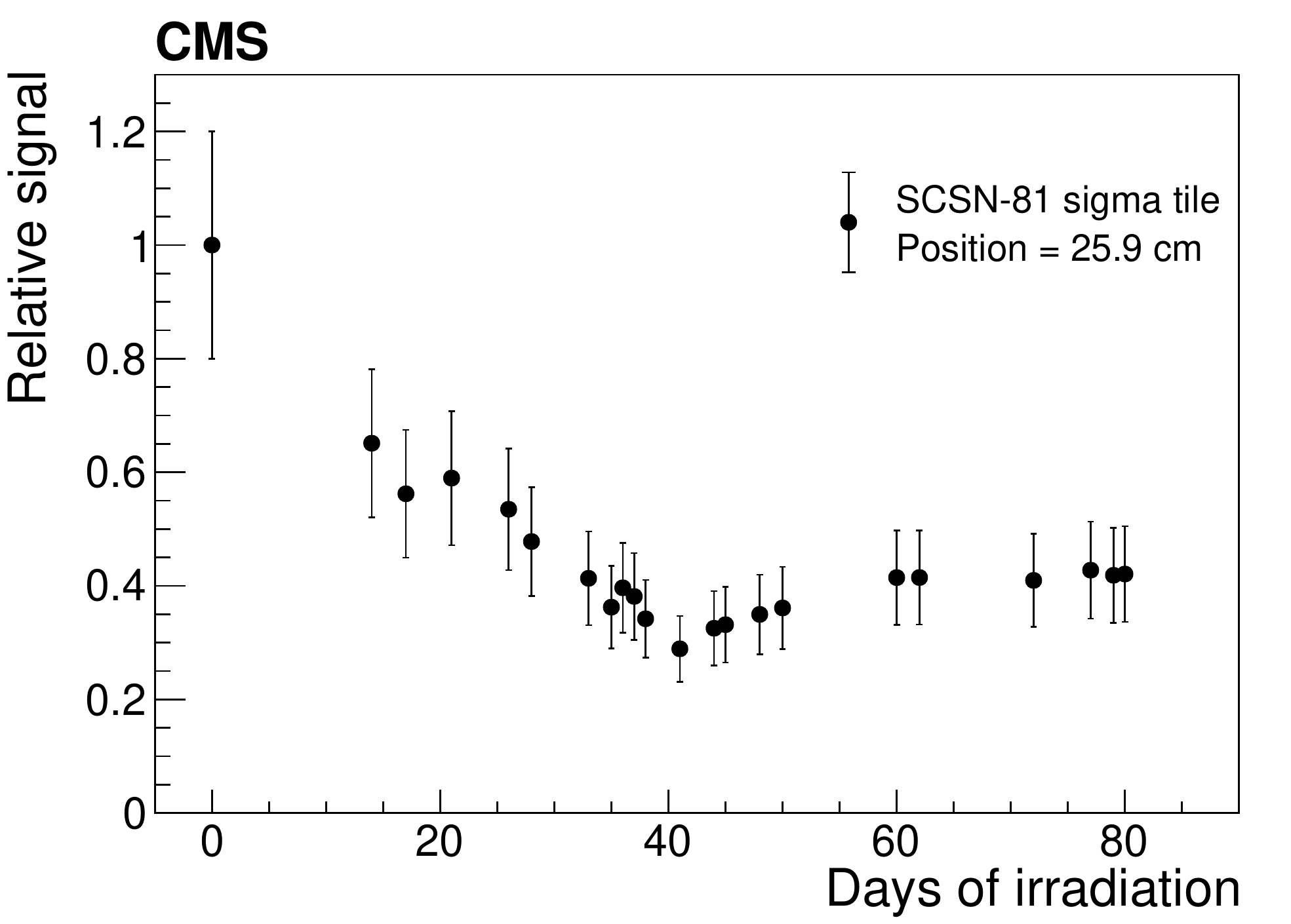}
\vskip 5mm
 \includegraphics[width=0.7\textwidth]{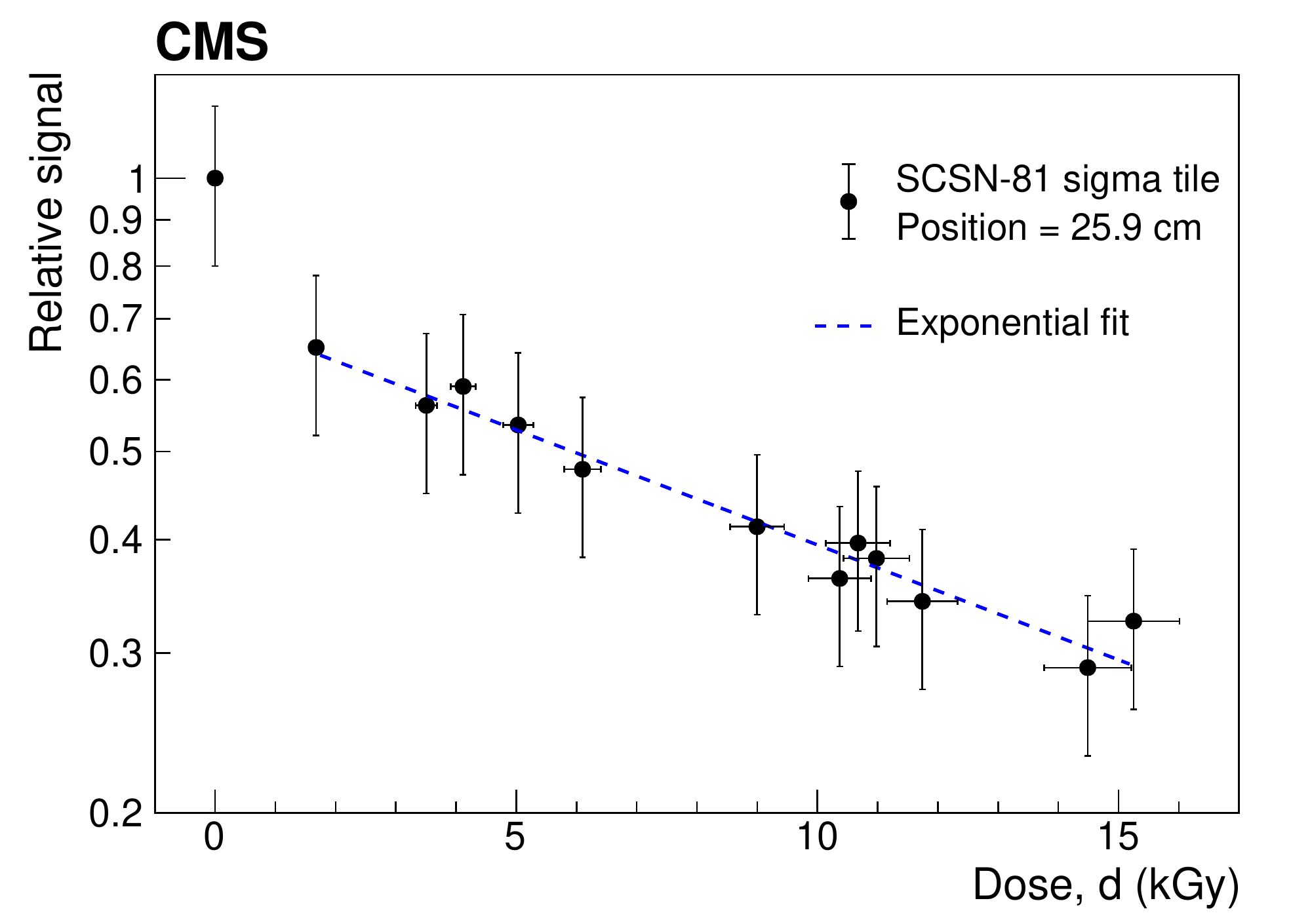}
 \caption{
Relative signal for an \SCSNeightone tile in the CRF radiation zone, plotted versus time (upper)
and versus received dose (lower), for $\dosrate = 42\unit{Gy/h}$.
}\label{fig:source-droprecovery}
\end{figure}
\begin{figure}[hbtp]\centering
 \includegraphics[width=0.8\textwidth]{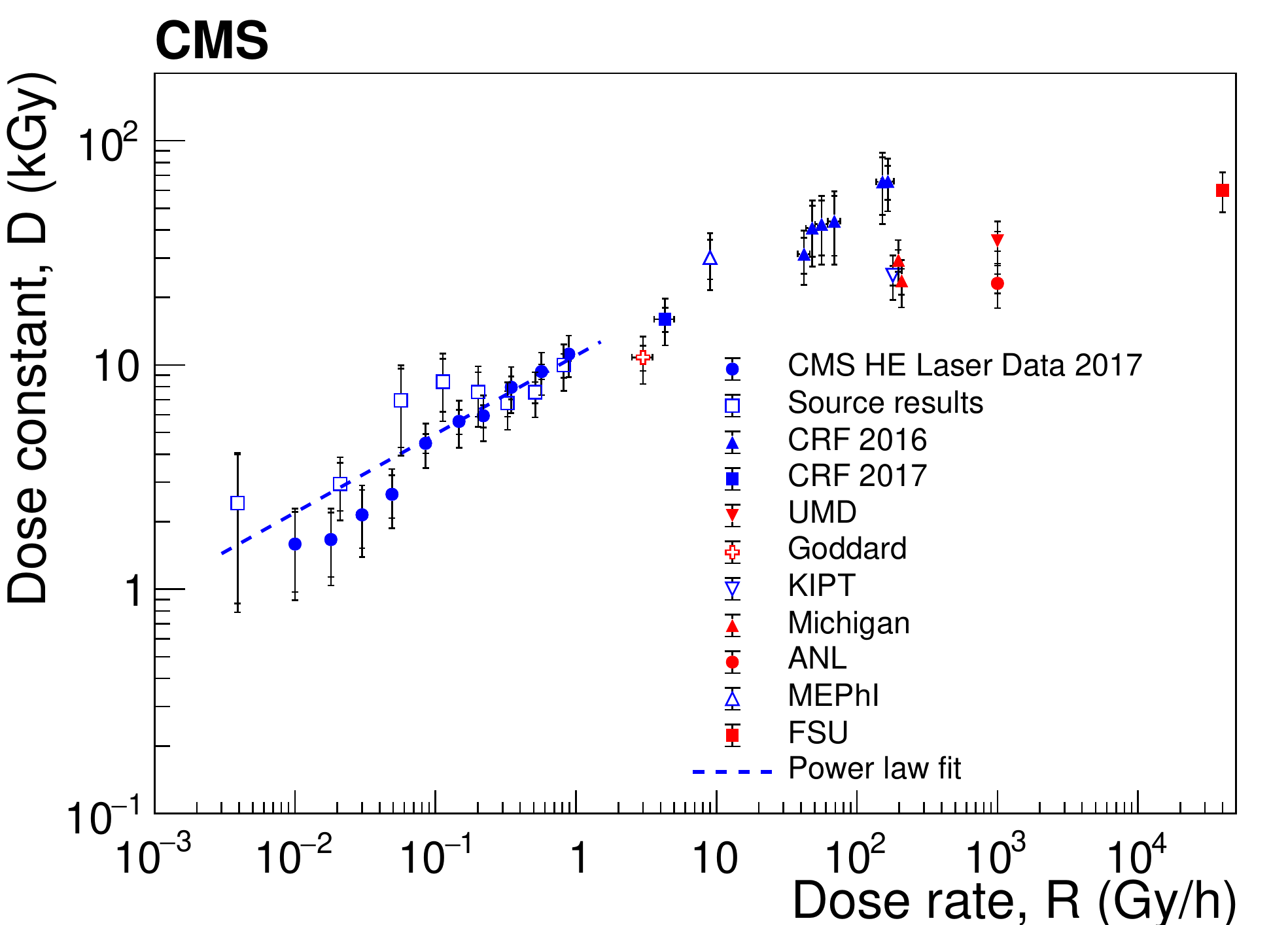}
 \caption{
Values of {\doseconst} versus {\dosrate} for high-{\dosrate} data taken with gamma irradiation
sources at KIPT, National Research Nuclear University MEPhI,
Goddard, Michigan, ANL, and UMD, an electron beam at FSU, and in the collider environment in the CRF
for \SCSNeightone tiles,
along with the results from the HE laser and source calibration data.
The statistical uncertainties are shown as the inner bars, and the outer bars include the systematic uncertainties
added in quadrature. The error bars on the irradiation data are dominated by systematic uncertainties.
}\label{fig:sourcedddot}
\end{figure}
\begin{figure}[hbtp]\centering
 \includegraphics[width=0.7\textwidth]{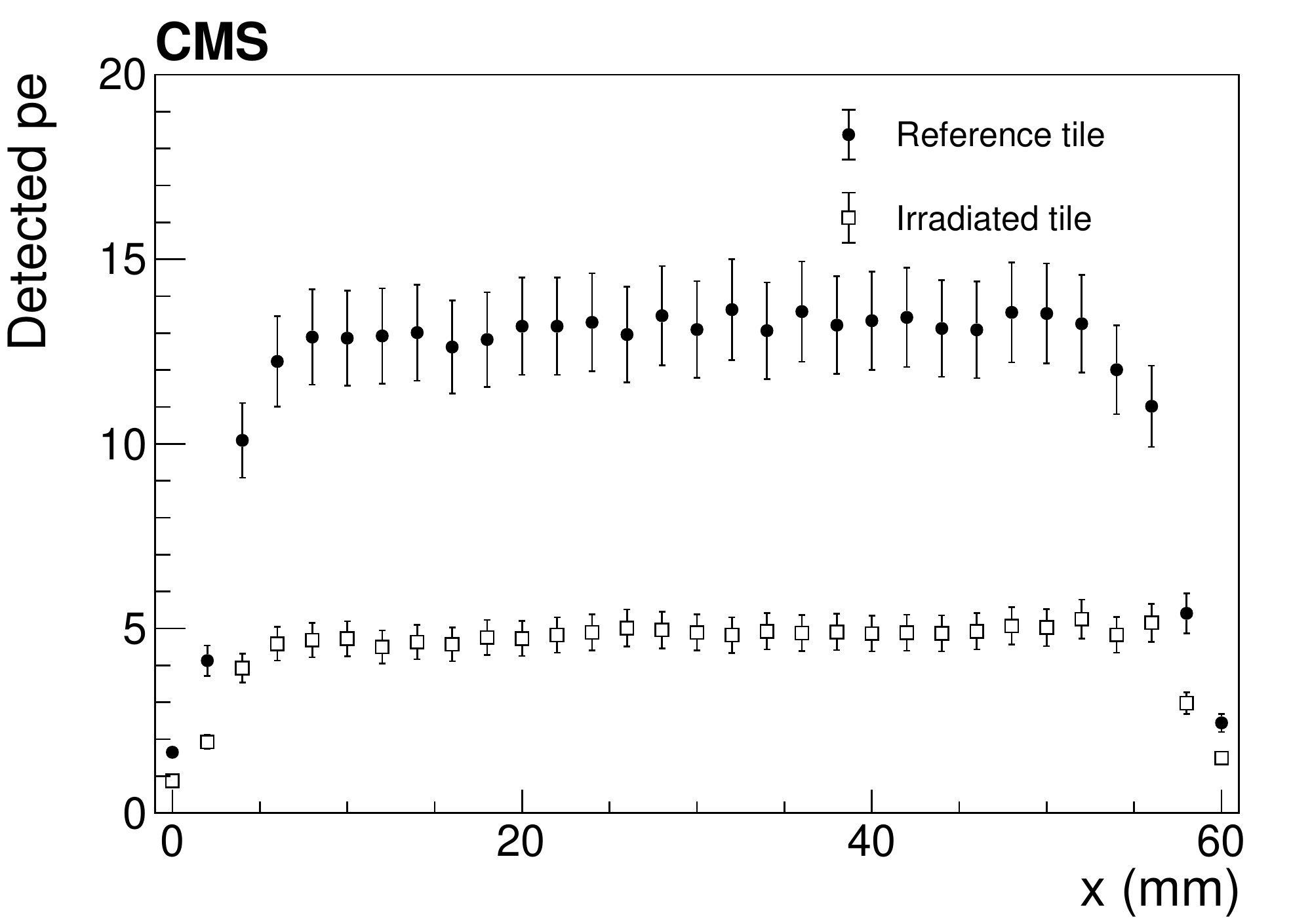}
 \caption{Number of detected photoelectrons
for an \SCSNeightone tile before and after an irradiation dose 
of  30\unit{kGy} at \dosrate of 9\unit{Gy/h},
as a function of the position of a radioactive-scan source along an axis
through the center of the tile and parallel to one of its sides.
The error bars are dominated by systematic uncertainty in normalization of the measurements;
statistical point uncertainties are~$<$2\%.
}\label{fig:transverse}
\end{figure}

For the source measurements, the  signal output of the samples was measured before and after irradiation
to calculate {\doseconst}.
The exact methods differ from study to study, but the general procedure involves the excitation of the
irradiated scintillator tile by particles (\eg, cosmic rays, or alpha or gamma particles
from a small, calibrated source placed in contact with the scintillator),
and the measurement of the signal output from the WLS fiber via either a photomultiplier tube or a \SIPM.

The remainder of the data were taken from samples irradiated in a region forward of CMS
called the CASTOR radiation facility (CRF).
These tiles were irradiated by particles originating from \pp collisions during the 2016 data-taking period.
They were located at radial distances from the beam line ranging from 11.8 to 25.9\unit{cm}.
The doses received by the CRF tiles in 2016 were determined based on film dosimetry measurements
and range from 15 to 60\unit{kGy}. An additional CRF-based measurement was performed during 2017,
using tiles at the radial distance from the beam of 43.2\unit{cm}, which received a dose of about 2.3\unit{kGy}.

For the CRF measurements, a laser calibration system was used to monitor the signal output of the tiles during the data taking.
As shown in Fig.~\ref{fig:source-droprecovery}, the signal loss
as a function of received dose appears to be more rapid in the initial stage of irradiation.
The tiles were remeasured in the laboratory after the CRF irradiation.
The results of these measurements indicate that
the initial drop seen in Fig.~\ref{fig:source-droprecovery} was caused by instrumental 
effects and not radiation damage.
The signal output follows an exponential decay for the remainder of the exposure.
There is some annealing after day 44, when the exposure ended.
The CRF data shown in Fig.~\ref{fig:sourcedddot} are corrected for the observed annealing.
Measurements of the tiles after removal from the CRF and replacement of the irradiated WLS fiber with a new
one indicate that about 20\% of the damage occured in the fiber.
The impact of radiation on reflectivity of Tyvek\texttrademark\  is
estimated by wrapping a single tile 
in the various Tyvek\texttrademark\ wrappers from the CRF samples exposed to different doses.
The light output of such sets was seen to decrease by about 0.2\% per 1 kGy of
the dose to the Tyvek\texttrademark\ wrappers. We conclude that the impact of Tyvek\texttrademark\ damage on the 
measurements of light output of the HE channels was negligible.

Figure~\ref{fig:sourcedddot} summarizes the results from the CRF and from electron beam 
and gamma source irradiations, along with the HE laser and source results.
We are not aware of other measurements of closely comparable tile-fiber systems
at the low dose rates seen by the HE scintillators.
For several orders of magnitude in \dosrate, \doseconst shows an apparent \dosrate-dependence.
The exact causes and mechanisms behind this effect remain to be understood.
In the next section, we compare the observed dependence to what is known about
dose-rate effects in plastic scintillators.

Tiles irradiated at gamma sources are also used to investigate the uniformity of the signal output after irradiation
and to check the dependence of {\doseconst} on the tile size.
A transverse scan of the signal output of a tile that received a total dose of
30\unit{kGy} at an {\dosrate} of 9\unit{Gy/h} is shown
in Fig.~\ref{fig:transverse}.  The number of  photoelectrons (pe) detected in scans prior to irradiation
is fairly independent of the source position.
The irradiated tile retains its uniformity after absorbing this large dose, implying that it is unlikely that optical light
attenuation is the major component of the observed signal loss.
Reference~\cite{Jivan_2015} came to similar conclusions based on Raman data, albeit for a PVT-based scintillator.

In addition, tiles with a thickness of 0.37\unit{cm} and sizes of $20\unit{cm} \times 20\unit{cm}$, 
$12\unit{cm} \times 9\unit{cm}$, and $5\unit{cm} \times 8\unit{cm}$
were irradiated at {\dosrate} of 1\unit{kGy/h} with doses of 1, 10, 20, 50, and 100\unit{kGy}.
The extracted values of $D$ are similar, to within $\pm$20\%.

We also investigated light propagation in tiles based on \GEANTfour~\cite{Agostinelli:2002hh} ray tracing.
Tile damage is simulated using the measured density of color centers. This study indicates that the effect of tile size
is expected to be small (at most 20\%).

\section{Discussion of dose-rate effects\label{sec:discussion}}

Because dose-rate effects have a significant impact on the performance
of scintillator-based detectors at hadron colliders,
in this section we review what is known of their origins.
Polymers are complex molecules, and their structure
depends on the details of their preparation
and the presence of additives such as antioxidants, while their behavior
depends in detail on their environment. Therefore, extrapolating
from measurements of a specific plastic in a specific environment
to another plastic and$/$or environment is difficult.
Measurements of new plastics and new environments will always be necessary.
However, existing theory facilitates a deeper understanding
of the results of our measurements.

Two well-studied~\cite{clough1,bolland1,bolland2,bateman,cunliffe,Wick1991472,Biagtan1996125}
sources of dose-rate effects in plastic scintillators
are related to oxygen, one involving the diffusion of oxygen into the plastic
during irradiation, and the other involving the rate of polymer oxidation in the areas containing oxygen.
Polymer oxidation can be either beneficial or detrimental, depending on the dose rate and
the details of the plastic preparation, the presence of additives such as antioxidants, and environment.
While the magnitude of polymer oxidation depends on such details,
theory gives us some guidance as to its dose-rate dependence.

As shown in the diagrams in Fig.~\ref{fig:color}, different kinds of termination, and thus permanent
color centers (see Section~\ref{sec:raddam}), are possible when oxygen is present.
Oxygen is highly reactive and polymer oxidation occurs
quickly after the production of the radicals~\cite{clough1,bolland1,bolland2,bateman,cunliffe,Wick1991472,Biagtan1996125}.
In this case, there is little of the temporary damage that is
indicative of radicals, and little to no annealing.
Since the final products involving oxygen tend to absorb
UV light, there can be considerable permanent damage that
results in what is called reduction of light output~\cite{cloughrad}
(see Section~\ref{sec:raddam}).
Temporary damage is larger without oxygen, as there is no oxygen to quickly bind to the radicals.
However, as the radicals slowly reform bonds, the resulting
stable structures sometimes have a small probability
to absorb visible light, reducing the plastic's absorption length.
Given the tension between these two competing effects, more experiments are needed to determine
the optimum atmosphere for different materials, dose rates, temperatures, and doses.
It is challenging to predict the optimal
amount of oxygen for a given value of \dosrate.

For a given plastic and environment, theory allows some
numerical extrapolation between different values of \dosrate.
At high enough \dosrate,  the density of radicals
produced is high enough that oxygen cannot diffuse into the plastic
fast enough to bind to and neutralize all the produced radicals,  and thus cannot penetrate beyond
a depth that depends on the dose rate~\cite{seguchi,cunliffe}.
The depth $z_0$ for oxygen diffusion into the plastic for a rectangular slab of plastic is~\cite{cunliffe}
\begin{linenomath}
\begin{equation}
z_0^2=\frac{2 \, M \, C_0}{\Upsilon \, \doserate}=\frac{2 \, M \, S \, P}{\Upsilon \, \doserate},
\label{eqn:z0}
\end{equation}
\end{linenomath}
where $M$ is the diffusion coefficient for oxygen, $C_0$ is the oxygen concentration on its edge,
$\Upsilon$ is the specific rate constant of active site formation, $S$ is the oxygen solubility, and
$P$ is the external oxygen pressure.
There is an abrupt transition between areas with and without oxygen.  The oxygen  concentration
in the oxidized regions is almost uniform~\cite{cloughPS}.
For PS tiles with a thickness of 4\mm, oxygen permeates the entire sample for
\dosrate below (roughly, depending on the plastic preparation and environment)
10\unit{Gy/h}~\cite{Wick1991472,cloughPS}; annealing should be small below this \dosrate.
For {\dosrate} above this value, polymer oxidation will occur
only in the region permeated by oxygen, contributing to an \dosrate dependence of the damage to the scintillator.

The second source of dose-rate effects is related to the
rate of polymer oxidation in regions with oxygen~\cite{clough1,bateman}.
The rate of polymer oxidation is~\cite{bolland1,clough1,clough2}
\begin{linenomath}
\begin{equation}
K(C(x,t))= -\frac{C_{1} \, C(x,t)}{1+C_{2} \, C(x,t)} ,
\label{eqn:diff2}
\end{equation}
\end{linenomath}
where $-K(C(x,t))$ is the rate at which oxygen is bound to the polymer, $x$ is the position relative
to the surface of the material where the rate is being measured,
and $C(x,t)$ is the concentration of oxygen.
The constants $C_1$ and $C_2$ depend on the kinematics of the chemical reactions.
The constant $C_1$ is related to polymer oxidation from radicals,
while $C_2$ is related to stable terminations of polymer oxidation.
The constant $C_1$ is proportional to the square root of {\dosrate} for bimolecular reactions
(leading to a dose-rate effect) and to {\dosrate} for unimolecular reactions (no dose-rate effect).

Another possible explanation for dose-rate effects involving oxygen for
acrylic scintillators (PMMA) is postulated in Ref.~\cite{Sirois}.  Radiation damage
in PMMA is generally larger, for the same dose, than in either PS or PVT.
The material produces more radicals and gas per dose
than PS or PVT and does not cross link~\cite{Wick1991472}.
The authors suggest that oxygen ions, produced by the radiation in the atmosphere surrounding the
material, may diffuse into the material and break polymer bonds,
and that the damage may be accentuated in the presence of UV light.
An irradiation at 0.1\unit{Gy/h} showed no damage when the samples were in a nitrogen atmosphere,
while damage was clearly seen for air and oxygen atmospheres.

According to Ref.~\cite{gillen}, dose-rate effects can also be caused by a change in the relative
amount of thermal- and radiation-induced damage.
At low {\dosrate}, damage due to thermal effects becomes more important.
Because thermal photons are of lower energy, they can only break
the lowest energy bonds, changing what types of radicals are formed.
This source of dose-rate effects is important when performing aging studies at high  temperature.

Other possible sources of dose-rate effects include damage to the fluors~\cite{Wick1991472},
damage to the fiber, presence of ozone~\cite{CloughOzone},
and an unknown mechanism observed in PS at high {\dosrate}
that is present at $22^\circ\unit{C}$ but not at $60^\circ\unit{C}$~\cite{cloughPS}.

Because dose-rate effects are seen in the HE tiles at $\dosrate  < 10\unit{Gy/h}$
when oxygen fully permeates the plastic,  the cause cannot be its penetration depth (see Eq.~\ref{eqn:z0}),
even though the power dependence close to 0.5 is suggestive.
The power dependence is in between that expected for unimolecular and bimolecular terminations of radicals
(see Eq.~\ref{eqn:diff2})~\cite{clough1,bolland1,bolland2,bateman,cunliffe,Wick1991472,Biagtan1996125}.
There is  a suggestion of a change of slope at a dose rate of 10\unit{Gy/h}, which, if real, could be
caused by different chemical processes in the regions with and without oxygen above this dose rate.

\section{Summary and conclusions\label{sec:summary}}

Radiation damage due to particles produced in \pp collisions at $\sqrt{s}=13\TeV$
in two types of plastic scintillator tiles in the CMS hadron endcap calorimeter
has been studied using data from several sources:
a laser calibration system, a movable radioactive source,
as well as hadrons and muons produced in \pp collisions.
Within the range of our measurements, the results from the various methods indicate that at
a fixed dose the damage to the scintillators increases with decreasing dose rate.
The dose-rate dependence is most accurately measured
by the laser system, with larger uncertainties in the other measurements.
The signal has an exponential decrease with dose characterized by dose constant $D$, which as a function
of dose rate \dosrate is compatible with a power law
with an exponent of about 0.4  for both PS and PVT-based tiles,
in between the values predicted by bimolecular and unimolecular
terminations of radicals~\cite{clough1,bolland1,bolland2,bateman,cunliffe,Wick1991472,Biagtan1996125}.
The PVT-based tiles indicate more damage than the PS-based tiles for the same exposure.
For $\doserate\approx 100\unit{Gy/h}$, approximately 20\% of the damage occurs in the fiber.
The results are compared to damage produced by irradiations with \CoSixty sources and by an electron beam.
At dose rates less than 10\unit{Gy/h},  relevant for future experiments at particle colliders,
where oxygen has saturated the plastic, the amount of damage does not depend on the particle type.

The parameters of the power-law fit are functions of the detector geometry, materials, ambient conditions, \etc
More studies are required to derive a general parametrization.
Nonetheless, fits such as these above have been used to predict the future behavior of the CMS hadron barrel
and endcap calorimeters~\cite{hcaltdr,hgcaltdr}.

Several aspects of the data-taking conditions in the
CMS detector give rise to systematic uncertainties that are difficult to estimate.
A set of identical tile + WLS fiber assemblies
subjected to varying dose-rate exposures in a temperature-controlled laboratory,
with careful monitoring throughout a year-long exposure, would allow for
a large reduction in the systematic uncertainties.
At high dose rates, the amount of damage has a considerable spread, possibly indicating
underestimated systematic uncertainties, motivating further studies to determine the underlying cause.
It would be interesting to have data over this wide range of dose rates separately for the fibers and for
the plastic tiles, to see their separate power dependencies.
Studies of tiles at low dose rates in an oxygen-free environment,
like a nitrogen atmosphere as suggested in Ref.~\cite{Sirois}, are needed
to test directly if the cause is dose-rate dependent polymer oxidation.
It would also be helpful to make measurements above 10\unit{Gy/h}
using a set of tiles made in a uniform way and irradiated at a known temperature.

Dose-rate effects can be large at low dose rates and should be measured for new tile systems.

\begin{acknowledgments}

We congratulate our colleagues in the CERN accelerator departments for the excellent performance of the LHC and thank the technical and administrative staffs at CERN and at other CMS institutes for their contributions to the success of the CMS effort. In addition, we gratefully acknowledge the computing centres and personnel of the Worldwide LHC Computing Grid for delivering so effectively the computing infrastructure essential to our analyses. Finally, we acknowledge the enduring support for the construction and operation of the LHC and the CMS detector provided by the following funding agencies: BMBWF and FWF (Austria); FNRS and FWO (Belgium); CNPq, CAPES, FAPERJ, FAPERGS, and FAPESP (Brazil); MES (Bulgaria); CERN; CAS, MoST, and NSFC (China); COLCIENCIAS (Colombia); MSES and CSF (Croatia); RPF (Cyprus); SENESCYT (Ecuador); MoER, ERC IUT, PUT and ERDF (Estonia); Academy of Finland, MEC, and HIP (Finland); CEA and CNRS/IN2P3 (France); BMBF, DFG, and HGF (Germany); GSRT (Greece); NKFIA (Hungary); DAE and DST (India); IPM (Iran); SFI (Ireland); INFN (Italy); MSIP and NRF (Republic of Korea); MES (Latvia); LAS (Lithuania); MOE and UM (Malaysia); BUAP, CINVESTAV, CONACYT, LNS, SEP, and UASLP-FAI (Mexico); MOS (Montenegro); MBIE (New Zealand); PAEC (Pakistan); MSHE and NSC (Poland); FCT (Portugal); JINR (Dubna); MON, RosAtom, RAS, RFBR, and NRC KI (Russia); MESTD (Serbia); SEIDI, CPAN, PCTI, and FEDER (Spain); MOSTR (Sri Lanka); Swiss Funding Agencies (Switzerland); MST (Taipei); ThEPCenter, IPST, STAR, and NSTDA (Thailand); TUBITAK and TAEK (Turkey); NASU (Ukraine); STFC (United Kingdom); DOE and NSF (USA).

\hyphenation{Rachada-pisek} Individuals have received support from the Marie-Curie programme and the European Research Council and Horizon 2020 Grant, contract Nos.\ 675440, 752730, and 765710 (European Union); the Leventis Foundation; the A.P.\ Sloan Foundation; the Alexander von Humboldt Foundation; the Belgian Federal Science Policy Office; the Fonds pour la Formation \`a la Recherche dans l'Industrie et dans l'Agriculture (FRIA-Belgium); the Agentschap voor Innovatie door Wetenschap en Technologie (IWT-Belgium); the F.R.S.-FNRS and FWO (Belgium) under the ``Excellence of Science -- EOS" -- be.h project n.\ 30820817; the Beijing Municipal Science \& Technology Commission, No. Z191100007219010; the Ministry of Education, Youth and Sports (MEYS) of the Czech Republic; the Deutsche Forschungsgemeinschaft (DFG) under Germany’s Excellence Strategy -- EXC 2121 ``Quantum Universe" -- 390833306; the Lend\"ulet (``Momentum") Programme and the J\'anos Bolyai Research Scholarship of the Hungarian Academy of Sciences, the New National Excellence Program \'UNKP, the NKFIA research grants 123842, 123959, 124845, 124850, 125105, 128713, 128786, and 129058 (Hungary); the Council of Science and Industrial Research, India; the HOMING PLUS programme of the Foundation for Polish Science, cofinanced from European Union, Regional Development Fund, the Mobility Plus programme of the Ministry of Science and Higher Education, the National Science Center (Poland), contracts Harmonia 2014/14/M/ST2/00428, Opus 2014/13/B/ST2/02543, 2014/15/B/ST2/03998, and 2015/19/B/ST2/02861, Sonata-bis 2012/07/E/ST2/01406; the National Priorities Research Program by Qatar National Research Fund; the Ministry of Science and Education, grant no. 14.W03.31.0026 (Russia); the Programa Estatal de Fomento de la Investigaci{\'o}n Cient{\'i}fica y T{\'e}cnica de Excelencia Mar\'{\i}a de Maeztu, grant MDM-2015-0509 and the Programa Severo Ochoa del Principado de Asturias; the Thalis and Aristeia programmes cofinanced by EU-ESF and the Greek NSRF; the Rachadapisek Sompot Fund for Postdoctoral Fellowship, Chulalongkorn University and the Chulalongkorn Academic into Its 2nd Century Project Advancement Project (Thailand); the Kavli Foundation; the Nvidia Corporation; the SuperMicro Corporation; the Welch Foundation, contract C-1845; and the Weston Havens Foundation (USA).
\end{acknowledgments}

\bibliography{auto_generated}
\cleardoublepage \appendix\section{The CMS Collaboration \label{app:collab}}\begin{sloppypar}\hyphenpenalty=5000\widowpenalty=500\clubpenalty=5000\vskip\cmsinstskip
\textbf{Yerevan Physics Institute, Yerevan, Armenia}\\*[0pt]
A.M.~Sirunyan$^{\textrm{\dag}}$, A.~Tumasyan
\vskip\cmsinstskip
\textbf{Institut f\"{u}r Hochenergiephysik, Wien, Austria}\\*[0pt]
W.~Adam, F.~Ambrogi, T.~Bergauer, J.~Brandstetter, M.~Dragicevic, J.~Er\"{o}, A.~Escalante~Del~Valle, M.~Flechl, R.~Fr\"{u}hwirth\cmsAuthorMark{1}, M.~Jeitler\cmsAuthorMark{1}, N.~Krammer, I.~Kr\"{a}tschmer, D.~Liko, T.~Madlener, I.~Mikulec, N.~Rad, J.~Schieck\cmsAuthorMark{1}, R.~Sch\"{o}fbeck, M.~Spanring, D.~Spitzbart, W.~Waltenberger, C.-E.~Wulz\cmsAuthorMark{1}, M.~Zarucki
% ...
\vskip\cmsinstskip
\dag: Deceased\\
1:  Also at Vienna University of Technology, Vienna, Austria\\
% ...
\end{sloppypar}
\end{document}